\documentclass[preprint2]{aastex631}
\usepackage{amsmath}

\shorttitle{The CEE Outcome}
\shortauthors{Ge et al.}

\begin{document}

\title{The Common Envelope Evolution Outcome---A Case Study on Hot Subdwarf B Stars}



\author[0000-0002-6398-0195]{Hongwei Ge}
\affiliation{Yunnan Observatories, Chinese Academy of Sciences,\\
396 YangFangWang, Guandu District, Kunming, 650216, China}
\affiliation{Institute of Astronomy, The Observatories, University of Cambridge,\\
	Madingley Road, Cambridge, CB3 0HA, UK}
\affiliation{Key Laboratory for Structure and Evolution of Celestial Objects, \\
	Chinese Academy of Sciences, P.O. Box 110, Kunming 650216, China}
\email{gehw@ynao.ac.cn}

\author[0000-0002-1556-9449]{Christopher A Tout}
\affiliation{Institute of Astronomy, The Observatories, University of Cambridge,\\
	Madingley Road, Cambridge, CB3 0HA, UK}
\email{cat@ast.cam.ac.uk}

\author[0000-0001-5284-8001]{Xuefei Chen}
\affiliation{Yunnan Observatories, Chinese Academy of Sciences,\\
	396 YangFangWang, Guandu District, Kunming, 650216, China}
\affiliation{Key Laboratory for Structure and Evolution of Celestial Objects, \\
	Chinese Academy of Sciences, P.O. Box 110, Kunming 650216, China}
\email{cxf@ynao.ac.cn}

\author[0000-0001-9331-0400]{Matthias U Kruckow}
\affiliation{Yunnan Observatories, Chinese Academy of Sciences,\\
	396 YangFangWang, Guandu District, Kunming, 650216, China}
\affiliation{Key Laboratory for Structure and Evolution of Celestial Objects, \\
	Chinese Academy of Sciences, P.O. Box 110, Kunming 650216, China}

\author{Hailiang Chen}
\affiliation{Yunnan Observatories, Chinese Academy of Sciences,\\
396 YangFangWang, Guandu District, Kunming, 650216, China}
\affiliation{Key Laboratory for Structure and Evolution of Celestial Objects, \\
	Chinese Academy of Sciences, P.O. Box 110, Kunming 650216, China}

\author[0000-0003-4265-7783]{Dengkai Jiang}
\affiliation{Yunnan Observatories, Chinese Academy of Sciences,\\
396 YangFangWang, Guandu District, Kunming, 650216, China}
\affiliation{Key Laboratory for Structure and Evolution of Celestial Objects, \\
	Chinese Academy of Sciences, P.O. Box 110, Kunming 650216, China}

\author[0000-0002-1421-4427]{Zhenwei Li}
\affiliation{Yunnan Observatories, Chinese Academy of Sciences,\\
	396 YangFangWang, Guandu District, Kunming, 650216, China}
\affiliation{Key Laboratory for Structure and Evolution of Celestial Objects, \\
	Chinese Academy of Sciences, P.O. Box 110, Kunming 650216, China}

\author[0000-0002-7909-4171]{Zhengwei Liu}
\affiliation{Yunnan Observatories, Chinese Academy of Sciences,\\
396 YangFangWang, Guandu District, Kunming, 650216, China}
\affiliation{Key Laboratory for Structure and Evolution of Celestial Objects, \\
	Chinese Academy of Sciences, P.O. Box 110, Kunming 650216, China}

\author[0000-0001-9204-7778]{Zhanwen Han}
\affiliation{Yunnan Observatories, Chinese Academy of Sciences,\\
396 YangFangWang, Guandu District, Kunming, 650216, China}
\affiliation{Key Laboratory for Structure and Evolution of Celestial Objects, \\
	Chinese Academy of Sciences, P.O. Box 110, Kunming 650216, China}

\email{zhanwenhan@ynao.ac.cn}

\begin{abstract}

Common envelope evolution (CEE) physics plays a fundamental role in the formation of binary systems, such as mergering stellar gravitational wave sources, pulsar binaries and type Ia supernovae. A precisely constrained CEE has become more important in the age of large surveys and gravitational wave detectors. We use an adiabatic mass loss model to explore how the total energy of the donor changes as a function of the remnant mass. This provides a more self-consistent way to calculate the binding energy of the donor. For comparison, we also calculate the binding energy through integrating the total energy from the core to the surface.  The outcome of CEE is constrained by total energy conservation at the point at which both component's radii shrink back within their Roche lobes. We apply our results to 142 hot subdwarf binaries. For shorter orbital period sdBs, the binding energy is highly consistent. For longer orbital period sdBs in our samples, the binding energy can differ by up to a factor of 2. The CE efficiency parameter $\beta_\mathrm{CE}$ becomes smaller than $\alpha_\mathrm{CE}$ for the final orbital period $\log_{10} P_{\mathrm{orb}}/\mathrm{d} > -0.5$.  We also find the mass ratios $\log_{10} q$ and CE efficiency parameters $\log_{10} \alpha_{\mathrm{CE}}$ and $\log_{10} \beta_{\mathrm{CE}}$ linearly correlate in sdBs, similarly to \citet{2011MNRAS.411.2277D} for post-AGB binaries. 

\end{abstract}



\section{Introduction} \label{sec:1}

Binary star evolution plays an essential role in the formation of almost all energetic and exotic objects in astrophysics. This hinges not only on the binary fraction being over half of all stellar systems \citep{2013ARA&A..51..269D,2017ApJS..230...15M} but also the existence of binary interactions, such as Roche-lobe overflow and common envelope evolution (CEE). CEE was initially proposed to explain the formation of short orbital period cataclysmic binaries \citep{1975PhDT.......165W,1976IAUS...73...75P}. Subsequently it has been found that CEE plays a fundamental role in the formation of binary systems with at least one compact component, a white dwarf, neutron star or black hole. These include mergering stellar gravitational wave sources \citep{2000ARA&A..38..113T,2001A&A...375..890N}, pulsar binaries \citep{1998A&A...332..173P}  and type Ia supernovae \citep{1984ApJ...277..355W}. 

\begin{figure}[ht!]
	\centering
	\includegraphics[scale=0.45]{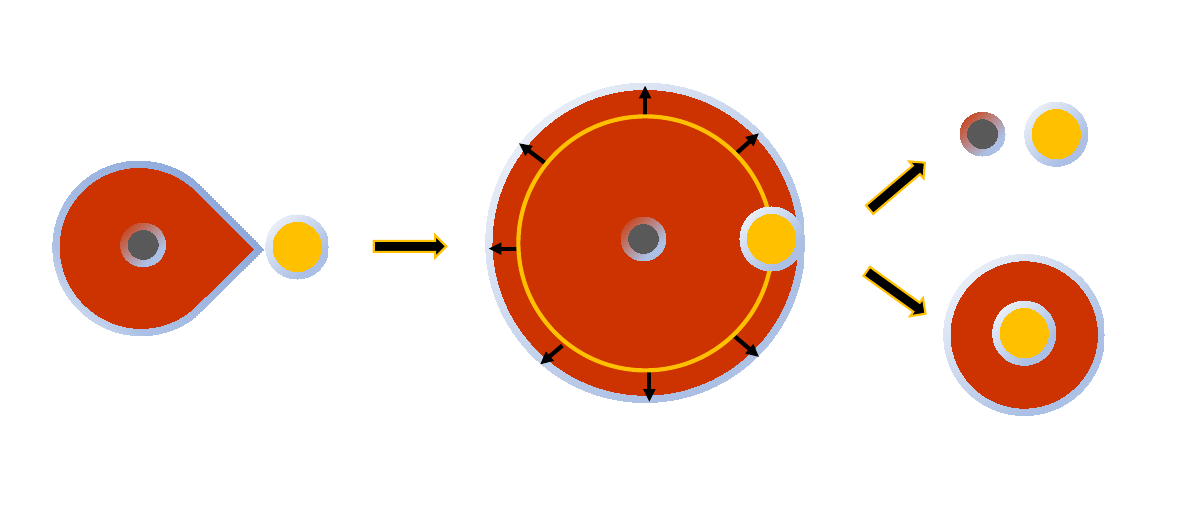}
	\caption{Common envelope evolution (CEE), not to scale. A giant unstably overflows on to a more compact companion. If the envelope can be ejected a close binary results. Otherwise the stars merge. \label{fig01}}
\end{figure}

A common envelope (CE) phase is believed to form when a giant donor star, with a deep convective envelope, transfers its mass to a less massive companion star \citep{1976IAUS...73...75P}. The outcome of a CE phase can be either the formation of a short orbital period compact binary in the case of a successful envelope ejection or the formation of a merged star in the case of a failure of envelope ejection (see Fig.\,\ref{fig01}). 

Generally speaking, there are two different approaches aimed at improving the CEE physics. The first is 3D hydrodynamic simulations. Significant progress has been made recently by addressing proper numerical methods and more precise physics. For examples, \citet{2012ApJ...746...74R} used both a piecewise parabolic method and an adaptive mesh refinement method to CE studies; \citet{2019IAUS..346..449R} considered radiation transport effects in CE simulations; \citet[and references therein]{2020A&A...644A..60S} examined the ionization energy contribution to the CE ejection process; \citet{2022MNRAS.tmp...68L} performed adiabatic simulations including both radiation energy and recombination energy. \citet{2022arXiv220509749G} simulated the CE process between a thermally pulsating asymptotic giant branch star and a low mass main-sequence (MS) star and found that (1) the inclusion of recombination energy resulting in wider separations;(2) thermal pulses can trigger CEs and lead to the prediction of a larger population of post-CE binaries. But a successful ejection of the whole envelope is still facing challenges. The second approach is a stellar evolution method. The standard procedure to predict the outcome of CEE is known as the energy formalism \citep{1984ApJ...277..355W, 1984ApJS...54..335I, 1988ApJ...329..764L}. This is based on the essential physics of total energy {\it conservation}. The initial and final total energy of the binary system $E_\mathrm{i}$ and $E_\mathrm{f}$ are assumed to be conserved.
\begin{equation}
	E_\mathrm{i} = E_\mathrm{f}
\end{equation}
with
\begin{equation}
	E_\mathrm{i} = E^\mathrm{i}_\mathrm{orb}+ E_\mathrm{1i}+ E_\mathrm{2i},
	\label{eq02}
\end{equation}
and
\begin{equation}
	E_\mathrm{f} = E^\mathrm{f}_\mathrm{orb}+E_\mathrm{1f}+ E_\mathrm{2f}.
	\label{eq03}
\end{equation}
The right parts of formulae~(\ref{eq02}) and~(\ref{eq03}) are the orbital energy of the binary system and the total self energy of the donor and accretor. Typically for a MS or white dwarf (WD) accretor $E_\mathrm{2}$ is assumed not to change. In this standard prescription a CE efficiency parameter  $\alpha_\mathrm{CE}$ is introduced as the fraction of the orbital energy change  $\Delta E_\mathrm{orb}$ that is used to overcome the binding energy  $E_\mathrm{bind}$. So that 
\begin{equation}
	\alpha_\mathrm{CE} \Delta E_\mathrm{orb} =\Delta E_\mathrm{1}= E_\mathrm{bind}.
	\label{eq04}
\end{equation}
This formula is popularly written as \citep{1984ApJ...277..355W,2008ASSL..352..233W, 1990ApJ...358..189D} 
\begin{equation}
	\alpha_\mathrm{CE} \left( -\frac{\mathrm{G}M_\mathrm{1i}M_\mathrm{2}}{2a_\mathrm{i}} +\frac{\mathrm{G}M_\mathrm{1f}M_\mathrm{2}}{2a_\mathrm{f}} \right) = \frac{\mathrm{G}M_\mathrm{1i}M_\mathrm{1e}}{\lambda R_\mathrm{1i} },
	\label{eq05}
\end{equation}
where $\mathrm{G}$ is Newton's gravitational constant, $M$ are the masses, $a$ are the semi-major axes, $R$ is the stellar radius and $\lambda$ is a dimensionless parameter that reflects the structure of the star. The 1 and 2 subscripts refer to the donor and accreter, respectively. The subscripts i and f refer to the initial and final state of the CE phase. $M_\mathrm{1e}=M_\mathrm{1i}-M_\mathrm{1f}$ is the mass of the common envelope.

The constraints on the onset, the outcome of CEE physics, and other basic knowledge of binary interactions can be used as physical inputs for binary population synthesis (BPS) studies. \citet{1987ApJ...318..794H} used polytropic stellar models to predict the critical initial mass ratios for dynamical timescale mass transfer, and \citet{1997A&A...327..620S} further extended this to non-conservatine cases. \citet{2003MNRAS.341..662C,2008MNRAS.387.1416C} used Cambridge STARS code and \citet{2015MNRAS.449.4415P} and \citet{2017MNRAS.465.2092P} used the MESA code to provide limits on the onset of CE process. \citet{2010ApJ...717..724G,2010Ap&SS.329..243G,2015ApJ...812...40G,2020ApJ...899..132G,2020ApJS..249....9G} built the adiabatic and thermal equilibrium mass-loss models and provided the critical initial mass ratios for dynamical timescale or thermal timescale mass transfer. Besides the BPS studies on low-mass binary stars \citep[eg.][]{2020MNRAS.499L.121L,2021MNRAS.501.1677H,2021ApJ...908..229L}, such studies for massive binary stars have increased recently. For examples, studies cover pulsating ultra-luminous X-ray sources \citep{2020A&A...642A.174M}, close double neutron stars sources \citep{2020PASA...37...38V,2020A&A...639A.123K,2021MNRAS.500.1380M}, neutron star binaries through accretion-induced collapse \citep{2020RAA....20..135W}, compact intermediate-mass black hole X-ray binaries \citep{2020ApJ...896..129C}, luminous red novae \citep{2021A&A...653A.134B}, rapidly rotating Be binaries \citep{2021MNRAS.502.3436E}, short-period massive binary stars \citep{2022A&A...659A..98S}, type II supernova progenitors \citep{2021A&A...645A...6Z} and black hole or neutron star binaries \citep[recent papers like][]{2021ApJ...920...81S,2022LRR....25....1M}. \citet{2021A&A...650A.107M} focuses on $30\,M_\odot$ models, while \citet{2016A&A...596A..58K} focuses on $80-88\,M_\odot$ models. Most of recent studies are driven by the detection of gravitational wave merges \citep{2016PhRvL.116m1103A,2017ApJ...848L..12A,2021PhRvX..11b1053A}. These observed binary objects need precisely constrained CEE physics.

However, the detailed physics of CEE outcome \citep[see the recent review by][]{2020cee..book.....I} is still far from well understood. Many authors provide useful constraints on an efficiency parameter $\alpha_\mathrm{CE}$ \citep[eg.][]{2010A&A...520A..86Z,2011MNRAS.411.2277D}, a structure parameter $\lambda$  or the binding energy \citep[eg.][]{2000A&A...360.1043D,2010ApJ...716..114X}. These studies are based on detailed calculations and provide the physical parameters as functions of stellar mass and evolutionary state. We provide here an alternative way to calculate the new binding energy from the difference between the initial and final total energy of the donor star \citep{2010ApJ...717..724G}. We investigate how the companion mass impacts the binding energy for a donor with the same mass and evolutionary state. We try to answer the question of whether there is a universal constant CE efficiency \citep{2020cee..book.....I} $\alpha_\mathrm{CE}$ through observed short-period hot subdwarf B stars (sdBs).

In Section \ref{sec:2}, we describe our method. In Section \ref{sec:3}, we introduce short orbital period sdBs from both theoretical and observational viewpoints. We present results and discussions of the CE efficiency parameters with binding energy derived from our standard formula and our new algorithm in Section \ref{sec:4}. We summarize and conclude on constraints on CEE outcomes in  Section \ref{sec:5}.

\section{Numerical Methods} \label{sec:2}

We use our code based on our adiabatic mass-loss model \citep{2010ApJ...717..724G,2010Ap&SS.329..243G,2015ApJ...812...40G,2020ApJ...899..132G} to investigate how the total energy varies as a function of the donor's remaining mass. The code replaces the luminosity and temperature equations with a frozen entropy profile to give the approximate response of a donor star undergoing very rapid mass transfer. The basic philosophy of the adiabatic mass-loss code is the same as the Cambridge STARS code \citep{1971MNRAS.151..351E,1972MNRAS.156..361E,1973MNRAS.163..279E} and a simplified version of it \citep{2004PASP..116..699P}. The input physics, such as the equation of state, opacity and nuclear reaction rates, is the same as in the STARS code \citep{1998MNRAS.298..525P}. We set the mixing-length parameter to be $\alpha = l/H_P = 2.0$ the ratio of mixing length $l$ to pressure scale height $H_P$ and the convective overshooting parameter $\delta_\mathrm{ov} = 0.12$ \citep{1998MNRAS.298..525P}.

\begin{figure}[ht!]
	\centering
	\includegraphics[scale=0.31]{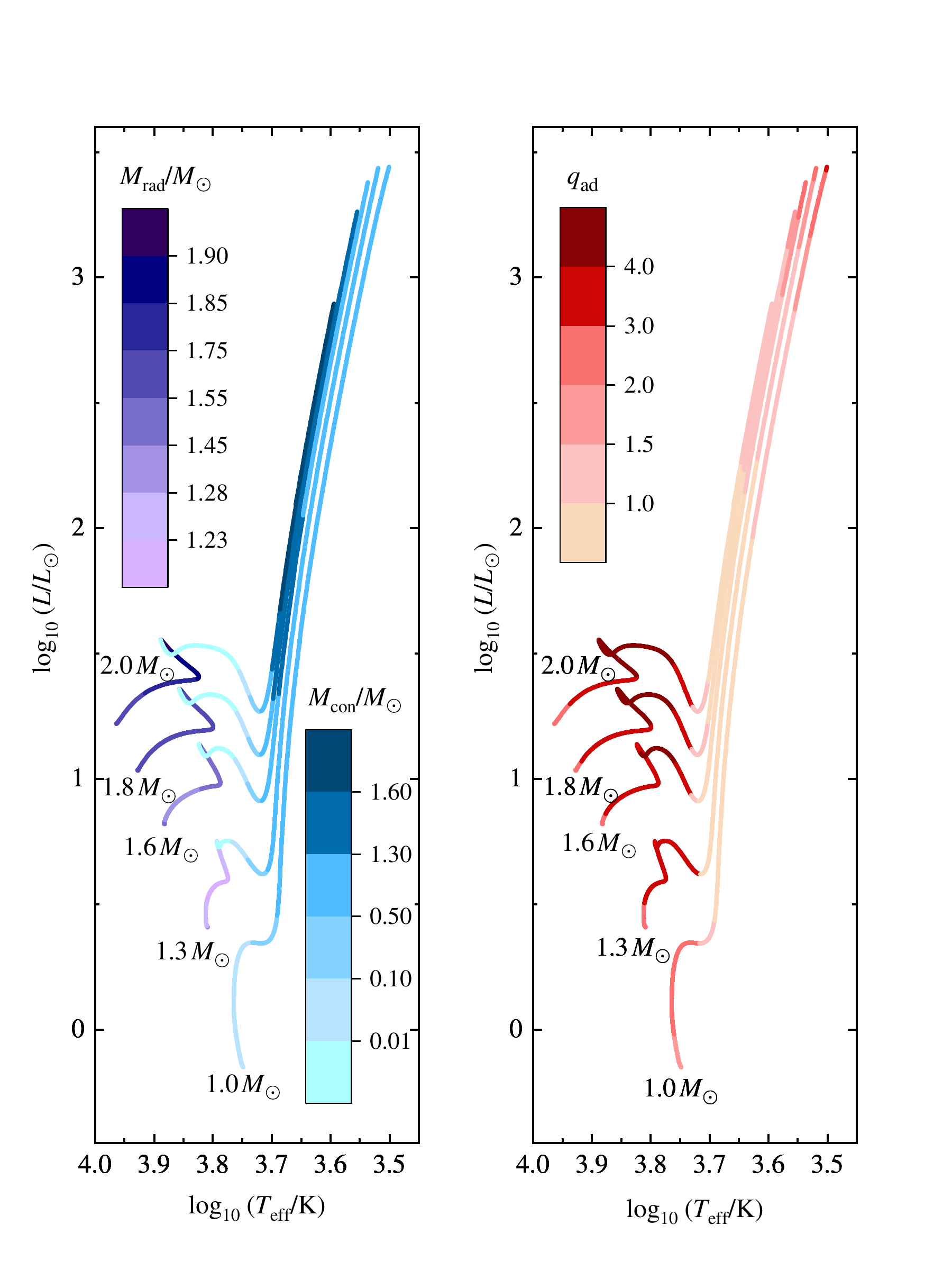}
	\caption{The Hertzsprung-Russell diagram of low-mass donor stars. The colors in the left panel indicate the mass of the convective or radiative envelope. The colors in the right panel show the critical initial mass ratio for dynamical timescale mass transfer \citep{2015ApJ...812...40G,2020ApJ...899..132G}. The tracks are for stars of $1.0$, $1.3$, $1.6$, $1.8$, and $2.0\,M_\odot$ from bottom to top. To avoid confusion, data ends at the tip of red giant branch (TRGB).
	\label{fig02}}
\end{figure}

We build a grid of initial stellar models with masses $1.0$, $1.3$, $1.6$, $1.8$ and $2.0\,M_\odot$ and metallicity $Z=0.02$ (see Fig.\,\ref{fig02}). The change of the total energy of the donor is calculated by \citep[see][]{2010ApJ...717..724G}
\begin{equation}
 \begin{aligned}
 	E_\mathrm{bind}= &\int_{0}^{M_{1 \mathrm{f}}}\left(-\frac{G m}{r}+U\right) d m~\\
    &-\int_{0}^{M_{1 \mathrm{i}}}\left(-\frac{G m}{r}+U\right) d m,
 \end{aligned}
 \label{eq06}
\end{equation}
where $m$ is the mass, $r$ is the radius and $U$ is its specific internal energy. Combining equations (\ref{eq04}) to (\ref{eq06}), we find the initial to final separation relation,
\begin{equation}
	\frac{a_{\mathrm{f}}}{a_{\mathrm{i}}}=\frac{M_{1 \mathrm{f}}}{M_{1 \mathrm{i}}}\left(1+\frac{2 a_{\mathrm{i}} \Delta E_{1}}{\beta_{\mathrm{CE}} G M_{2} M_{1 \mathrm{i}}}\right)^{-1}.
	\label{eq07}
\end{equation}
Here, we define the CE efficiency parameter as $\beta_\mathrm{CE}$ to indicate that the binding energy is calculated by equation (\ref{eq06}) which considers the response and the redistribution of the core and the thin envelope. This can differ somewhat from that of the initial stellar model, particularly for more massive stars with a non-degenerate core \citep{2010ApJ...717..724G,2010ApJ...719L..28D}. For comparison, we also calculate the binding energy from
\begin{equation}
	 E_\mathrm{bind}^{'}= -\int_{M_ \mathrm{He}}^{M_\mathrm{1i}}\left(-\frac{G m}{r}+U\right) d m,
	\label{eq08}
\end{equation}
and $M_\mathrm{He}$ is the helium core mass where the mass fraction of hydrogen is $0.1$. We also transfer this binding energy to the corresponding structure parameter $\lambda$, see table \ref{tab01}, through $E_\mathrm{bind}^{'} = GM_\mathrm{1i}M_\mathrm{1e}/\lambda R_\mathrm{1i}$. In this case, the corresponding CE efficiency parameter $\alpha_\mathrm{CE}$ is the same as in formula (\ref{eq05}).

\begin{deluxetable*}{cccccc}
	\tablenum{1}
	\tablecaption{Parameters about three TRGB stars
	\label{tab01}}
	\tablewidth{0pt}
	\tablehead{
		\colhead{$M_\mathrm{1i}$} & \colhead{$M_\mathrm{He}$} & \colhead{$\log_{10} R_\mathrm{1i}$} & \colhead{$Z$} & \colhead{$E_\mathrm{bind}^{'}$} & \colhead{$\lambda$} \\
			\colhead{$M_\odot$} & \colhead{$M_\odot$} & \colhead{$R_\odot$}&\colhead{ } &  \colhead{erg} & \colhead{ } 
	}
	\startdata
	1.0 & 0.4620 & 2.2425 & 0.02 & $2.548\times 10^{46}$ & 0.4580 \\
	1.3 & 0.4615 & 2.2035 & 0.02 & $3.029\times 10^{46}$ & 0.8540 \\
	1.6 & 0.4559 & 2.1394 & 0.02 & $4.235\times 10^{46}$ & 1.1895 \\
	\enddata

\end{deluxetable*}

The initial separation $a_\mathrm{i}$ can be found, from the donor star's mass $M_\mathrm{1i}$, radius $R_\mathrm{1i}=R_\mathrm{L}$ (requiring the donor to fill its Roche lobe at the onset of the CE) with companion mass $M_\mathrm{2}$, to be
\begin{equation}
	a_\mathrm{i} = \frac{R_\mathrm{1i}}{r_\mathrm{L}(M_\mathrm{1i}/M_\mathrm{2})},
	\label{eq09}
\end{equation}
where $R_\mathrm{1i}$ is the radius of the giant donor when it first fills its Roche lobe. For sdB stars we take this to be the radius at the tip of the red giant branch (TRGB)\footnote{If a progenitor overfills its Roche lobe near the TRGB and suffers a CE ejection, the remnant core can trigger a delayed helium flash and finally form a sdB (\citealt{2002MNRAS.336..449H}). So, we set $R_\mathrm{1i} = R_\mathrm{TRGB}$.}, and, by Eggleton's approximation for Roche-lobe radius \citep{1983ApJ...268..368E},
\begin{equation}
	r_{\rm L}(q) = \frac{0.49 q^{2/3}}{0.6 q^{2/3} + \ln(1+q^{1/3})} = \frac{R_{\rm L}}{a}\ .
	\label{eq10}
\end{equation}
The final separation $a_\mathrm{f}$ can be determined from an observed object's orbital period if we assume the period and semi-major axis did not change between the end of the CE and the observation, 
\begin{equation}
	P_\mathrm{orb}^2 = \frac{4 \pi^2 a_\mathrm{f}^3}{G(M_\mathrm{1f}+M_2)}.
	\label{eq11}
\end{equation}

The outcome of CEE can be self-consistently constrained by total energy conservation (equation \ref{eq07}) at the point at which both component's radii shrink back within their Roche lobes, $R_\mathrm{1f} \leq R_\mathrm{1,L_1}$ and $R_\mathrm{2} \leq R_\mathrm{2,L_1}$ \citep[see also][]{2010ApJ...717..724G}.

\section{Short Orbital Period \lowercase{sd}B\lowercase{s}} \label{sec:3}

Subdwarf B stars, sdBs, are core helium-burning stars with masses around $0.5\,M_\odot$ and a thin hydrogen-rich envelope (less than $0.02\,M_\odot$). They are hot and their effective temperature is between 20,000 and 40,000\,K \citep{2009ARA&A..47..211H}. So sdBs emit strong ultraviolet (UV) emission and this makes them the major source of the UV-upturn in the elliptical galaxies \citep{1991ApJ...382L..69F,2007MNRAS.380.1098H}. About half of sdBs are in close binaries with a WD or low-mass MS companion star \citep{2009ARA&A..47..211H}. The most favoured formation channels for sdBs are the CE ejection channel, the stable Roche lobe overflow (RLOF) channel and the merging double helium white dwarf channel \citep{2002MNRAS.336..449H,2003MNRAS.341..669H}. The merging channel produces isolated sdBs and the first two provide sdBs in binaries. The CEE with a sub-stellar companion can also form an isolated sdB star \citep{1998AJ....116.1308S,2020A&A...642A..97K}. Besides their importance in binary evolution, sdBs are also useful for asteroseismology \citep[e.g.][]{2003ApJ...597..518F}, cosmology \citep{2009A&A...493.1081J,2018RAA....18...49W} and gravitational wave sources studies \citep{2018A&A...618A..14W}.

\begin{figure}[ht!]
	\centering
	\includegraphics[scale=0.31]{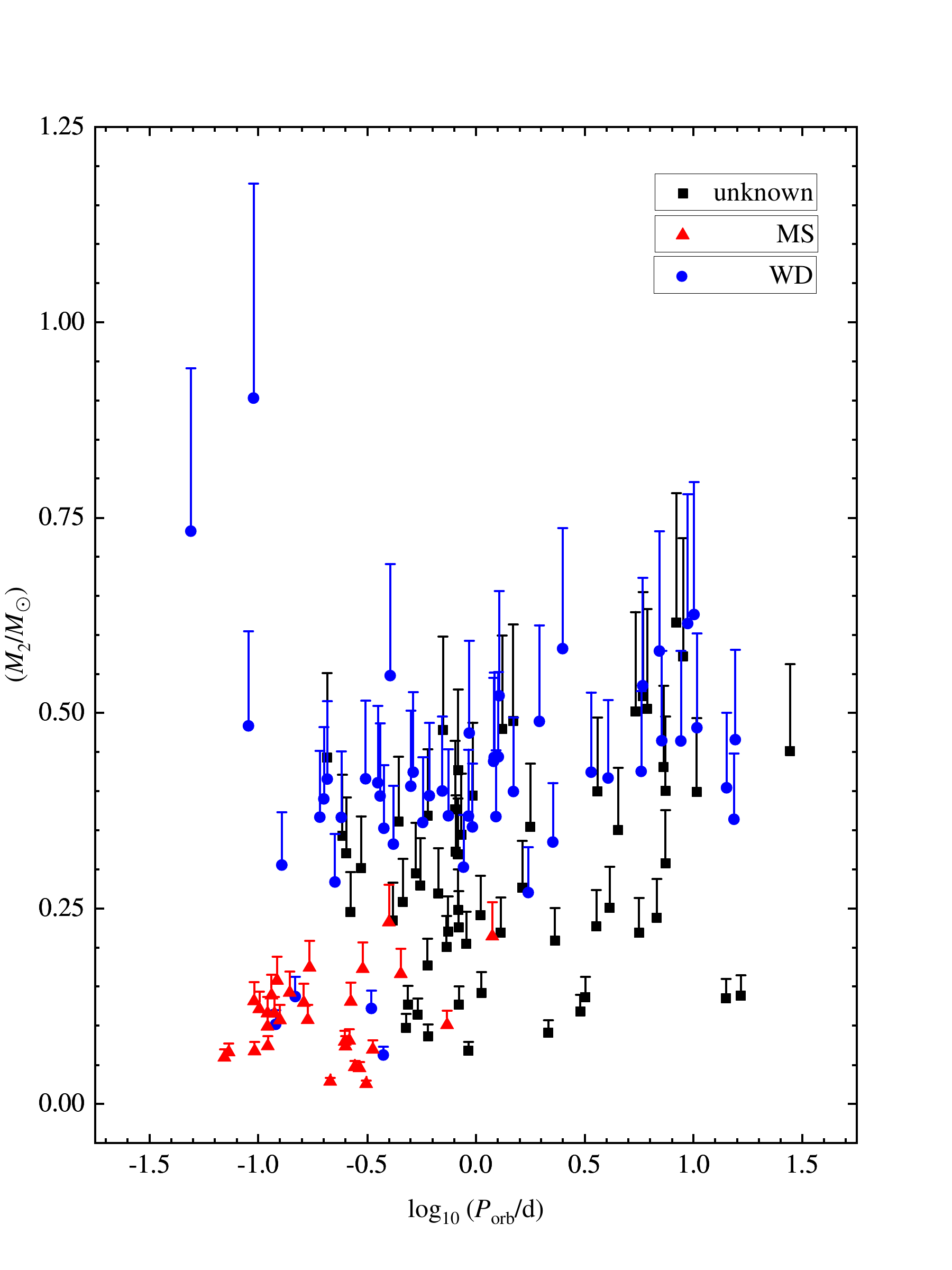}
	\caption{The orbital periods against the companion masses of 142 observed hot subdwarf B stars (sdBs). As described by \citet{2015A&A...576A..44K}, different symbols indicate the minimum companion masses, assuming the sdB mass  $M_\mathrm{sdB}=0.47\,M_\odot$ and the inclination angle $ i= 90^\circ$. A mean companion mass, the upper error bar, can be found assuming an inclination angle $i=60^\circ$. Black squares, red triangles, and blue circles show the type of the companion star, unknown, main sequence (MS), or white dwarf (WD), respectively. 
		\label{fig03}}
\end{figure}

Recently, many sdBs have been found in survey projects, such as Gaia \citep{2019A&A...621A..38G} and LAMOST \citep{,2021ApJS..256...28L}. However, a strict constraint on the orbital parameters and masses of both components is still lacking. So we use sdB examples provided by \citet{2015A&A...576A..44K} to carry out our study. Through detailed stellar evolution studies, \citet{2002MNRAS.336..449H} and \citet{2021MNRAS.505.3514Z} find that the low-mass progenitor models of short orbital period sdBs are located almost at the TRGB in order to trigger a helium flash. Furthermore, the CE ejection scenario is the only likely formation channel for these short-period sdB binaries below 10\,d \citep{2001MNRAS.326.1391M}. Hence, the simplicity of these sdBs' progenitors makes them the perfect objects to constrain CEE physics.

\citet{2015A&A...576A..44K} studied 142 short-period sdB binaries, including 12 new systems. Assumming a canonical mass for the sdB $M_\mathrm{sdB}=0.47\,M_\odot$ \citep{2012A&A...539A..12F} and inclination angles of $i = 90^\circ \mathrm{and}~ 60^\circ$, the minimum and mean mass of the companion can be determined by solving the equation for the mass function (Fig.\,\ref{fig03}). \citet{2015A&A...576A..44K} find that a peak around $0.1\,M_\odot$ corresponds to the low-mass MS companions and another peak around $0.4\,M_\odot$ corresponds to the WD companions (see their Fig.\,8). Again, these short-period sdB binaries are most likely formed from CEE and we use them to constrain its physics in the next section.

\section{Results and Discussions} \label{sec:4}

We present CE efficiency parameters as a function of the orbital period for 142 sdB binaries. Three different TRGB donor stars are explored. Their binding energies are calculated by either integrating from the core to the surface of the initial model when CEE begins $E^{'}_\mathrm{bind}$ (equation \ref{eq08}) or the change of total energy during CEE $E_\mathrm{bind}$ (equation \ref{eq06}). The corresponding CE efficiency parameters are writen as $\alpha_\mathrm{CE}$ and $\beta_\mathrm{CE}$, respectively. For shorter orbital period sdBs, the binding energy is almost the same and we find the difference between $\alpha_\mathrm{CE}$ and $\beta_\mathrm{CE}$ is small. For longer orbital period sdBs in our samples we have found that the binding energy can differ by up to a factor of 2. The CE efficiency parameter $\beta_{\mathrm{CE}}$ becomes smaller than $\alpha_\mathrm{CE}$ for final orbital periods $\log_{10} P_{\mathrm{orb}}/\mathrm{d} > -0.5$. A shallower slope for $\log_{10} \beta_{\mathrm{CE}}$ with $\log_{10} P_{\mathrm{orb}}$ than that for $\log_{10} \alpha_{\mathrm{CE}}$ is found for sdBs with all types of companion.

\subsection{1.0 $M_\odot$ TRGB Donor Star}

\begin{figure}[ht!]
	\centering
	\includegraphics[scale=0.31]{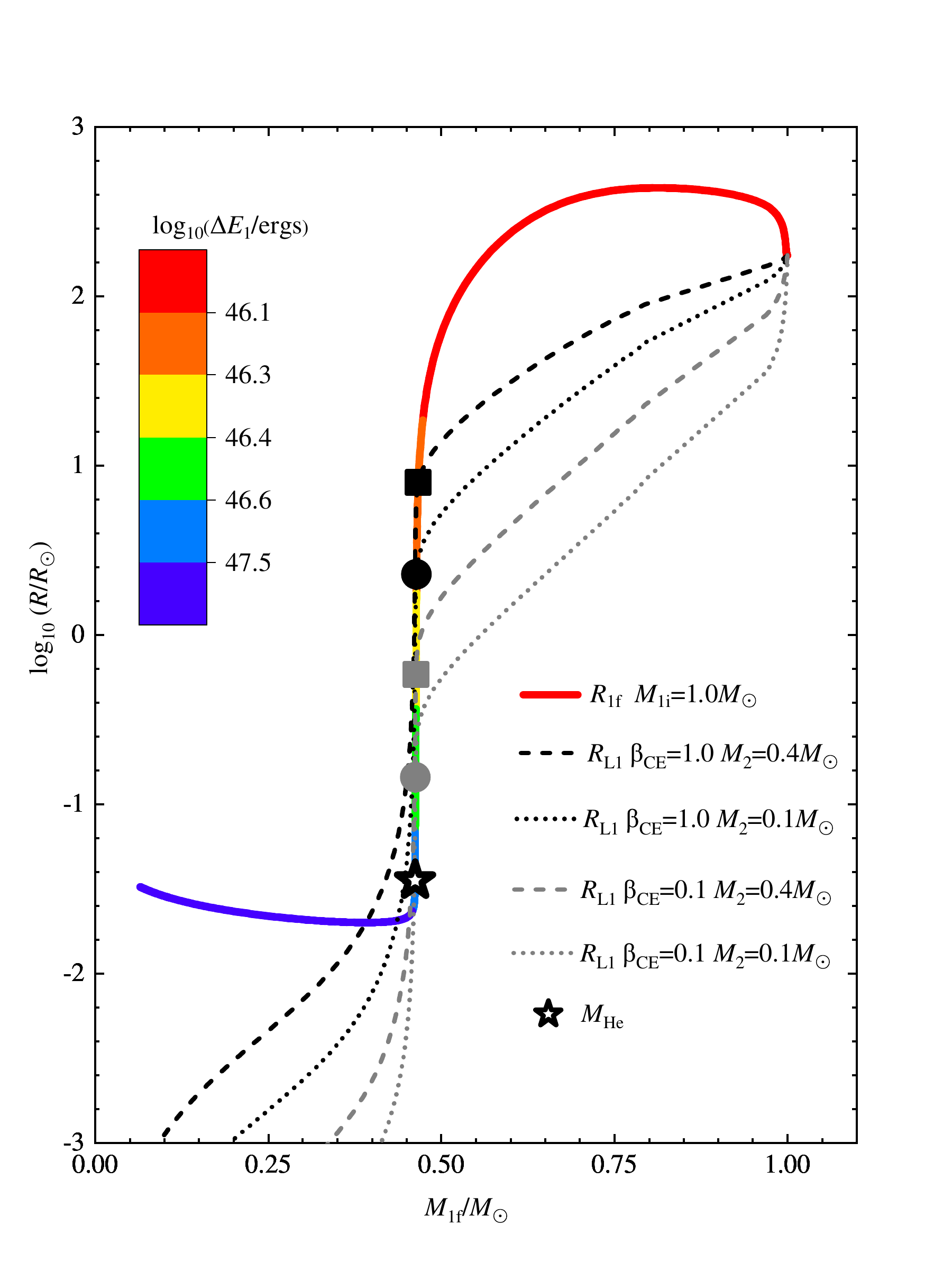}
	\caption{The remnant radius (colored line) and Roche-lobe radius (dashed and dotted lines) as a function of mass. The initial model is a $1.0\,M_\odot$ star at the TRGB. Dashed and dotted lines (solving equations \ref{eq07} to \ref{eq11} of this paper and equation 1 of \citealt{2015A&A...576A..44K}) correspond to companion masses of $0.4\,M_\odot$ and $0.1\,M_\odot$. Black and gray indicate efficiency parameters $\beta_\mathrm{CE}$ of 1.0 and 0.1. Filled squares and filled circles locate the outcome of the CEE with the given efficiency parameter $\beta_\mathrm{CE}$ and companion mass $M_2$. The pentagram is the helium core mass $M_\mathrm{He}$ where the mass fraction of hydrogen is 0.1.
		\label{fig04}}
\end{figure}

\begin{figure}[ht!]
	\centering
	\includegraphics[scale=0.31]{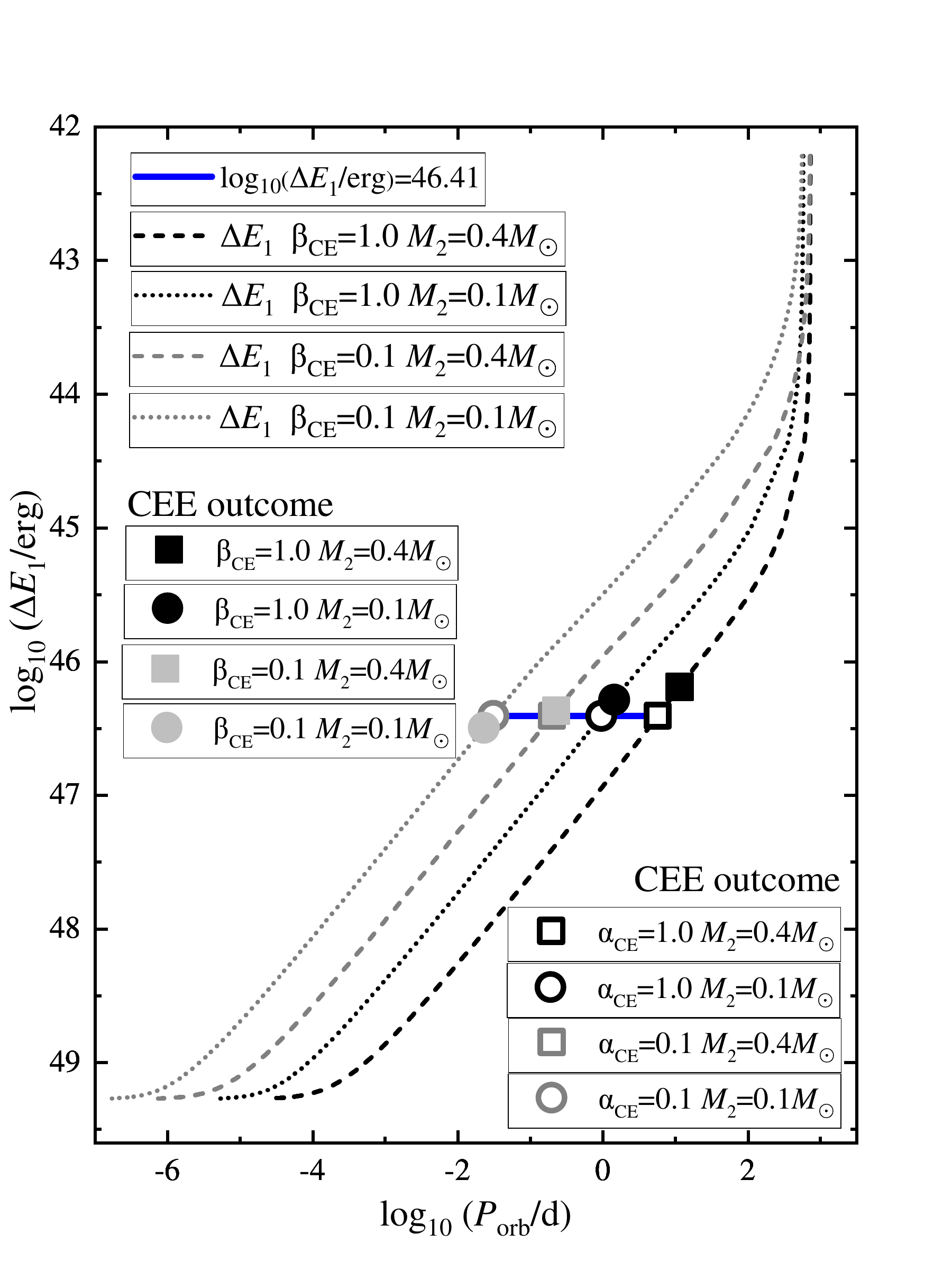}
	\caption{The binding energy as a function of orbital period. The initial model is a $1.0\,M_\odot$ star at the TRGB. Dashed and dotted lines (solving equations \ref{eq07} to \ref{eq11} of this paper and equation 1 of \citealt{2015A&A...576A..44K}) correspond to companion masses of $0.4\,M_\odot$ and $0.1\,M_\odot$. Black and gray indicate efficiency parameters $\beta_\mathrm{CE}$ of 1.0 and 0.1. Filled ($\beta_\mathrm{CE}$) and open ($\alpha_\mathrm{CE}$) symbols locate the outcome of the CEE with the given efficiency parameter and companion mass. Filled symbols correspond to each other between Figs.\,\ref{fig04} and \ref{fig04b}. The blue line is the binding energy from the right handside of equation (\ref{eq05}) for $\lambda = 0.458$.
		\label{fig04b}}
\end{figure}

\begin{deluxetable*}{ccccccc}
	\tablenum{2}
	\tablecaption{Fitting parameters of different type companions
		\label{tab02}}
	\tablewidth{0pt}
	\tablehead{
		\colhead{type} & \colhead{efficiency} & \colhead{intercept value} & \colhead{standard error} & \colhead{slope value} & \colhead{standard error} & \colhead{adj. R-Square} \\
	}
	\startdata
	MS      & $\log_{10} \alpha_{\mathrm{CE}}$ & -0.02515 & 0.11047 & 0.58893 & 0.14319 & 0.35437 \\
	MS      & $\log_{10} \beta_{\mathrm{CE}}$ & -0.15583 & 0.10001 & 0.47214 & 0.12962 & 0.29725 \\
	WD      & $\log_{10} \alpha_{\mathrm{CE}}$ & -0.49692 & 0.02341 & 0.61430 & 0.03590 & 0.85120 \\
    WD      & $\log_{10} \beta_{\mathrm{CE}}$ & -0.60204 & 0.02289 & 0.50816 & 0.03511 & 0.80345 \\
    unknown & $\log_{10} \alpha_{\mathrm{CE}}$ & -0.34292 & 0.02740 & 0.62956 & 0.04982 & 0.72893 \\
    unknown & $\log_{10} \beta_{\mathrm{CE}}$ & -0.45605 & 0.02532 & 0.51224 & 0.04603 & 0.67553 \\
	\enddata
\tablecomments{1. Standard error was scaled with square root of reduced Chi-square.\\
2. R-Square equals $1 - \sum_{i=1}^n (y_i -f_i)^2/\sum_{i=1}^n (y_i-\overline{y})^2$ where n is the total number of these 142 sdBs, $y_i$ is the logrithmic of CE efficiency, $f_i$ is the fitted value,  $\overline{y}$ is the mean of $y_i$.	}
\end{deluxetable*}

\begin{figure}[ht!]
	\centering
	\includegraphics[scale=0.3]{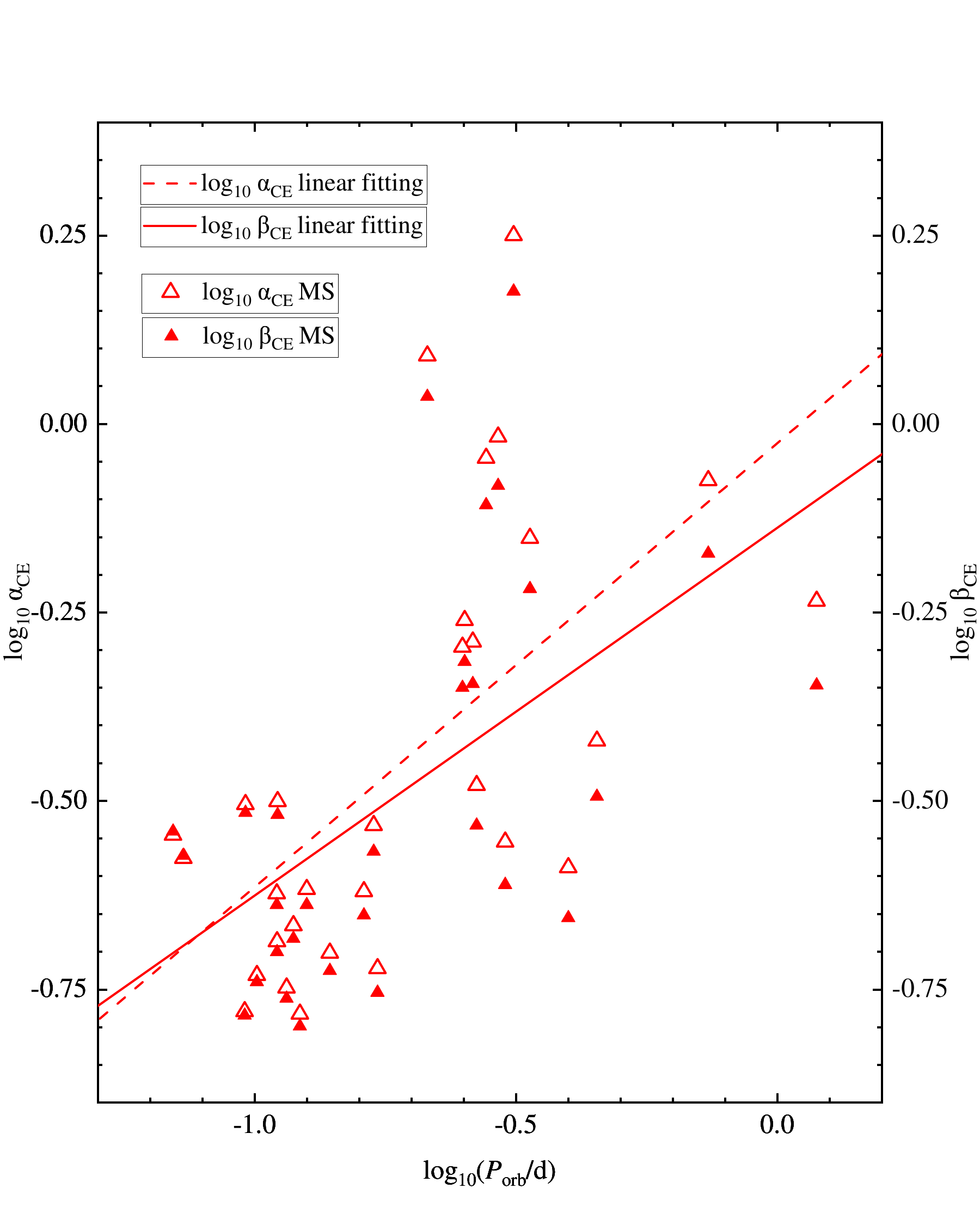}
	\caption{The CE efficiency parameters $\alpha_\mathrm{CE}$ (open symbols) and $\beta_\mathrm{CE}$ (filled symbols) against the orbital period of observed sdBs with a MS companion. The initial model is a $1.0\,M_\odot$ TRGB star (other initial models are considered in Figs \ref{fig09} to \ref{fig11}). The companion mass $M_2$ is solved with an inclination angle of $i = 90^\circ$. The dashed and solid lines are a linear least-square fits to the data. 
	\label{fig05}}
\end{figure}

\begin{figure}[ht!]
	\centering
	\includegraphics[scale=0.3]{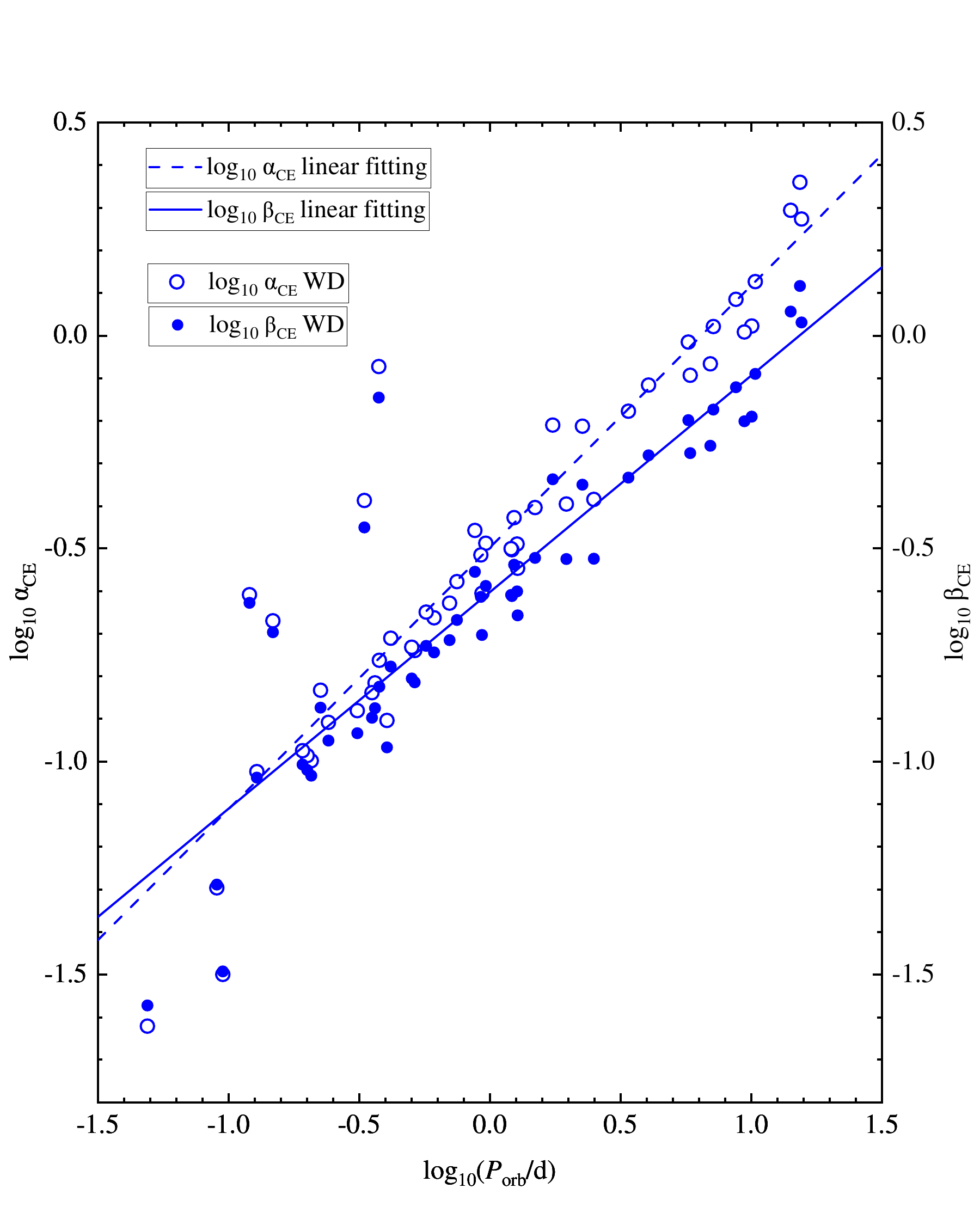}
	\caption{Similiar to Fig.\,\ref{fig05} but for sdBs with a WD companion.
		\label{fig06}}
\end{figure}

\begin{figure}[ht!]
	\centering
	\includegraphics[scale=0.3]{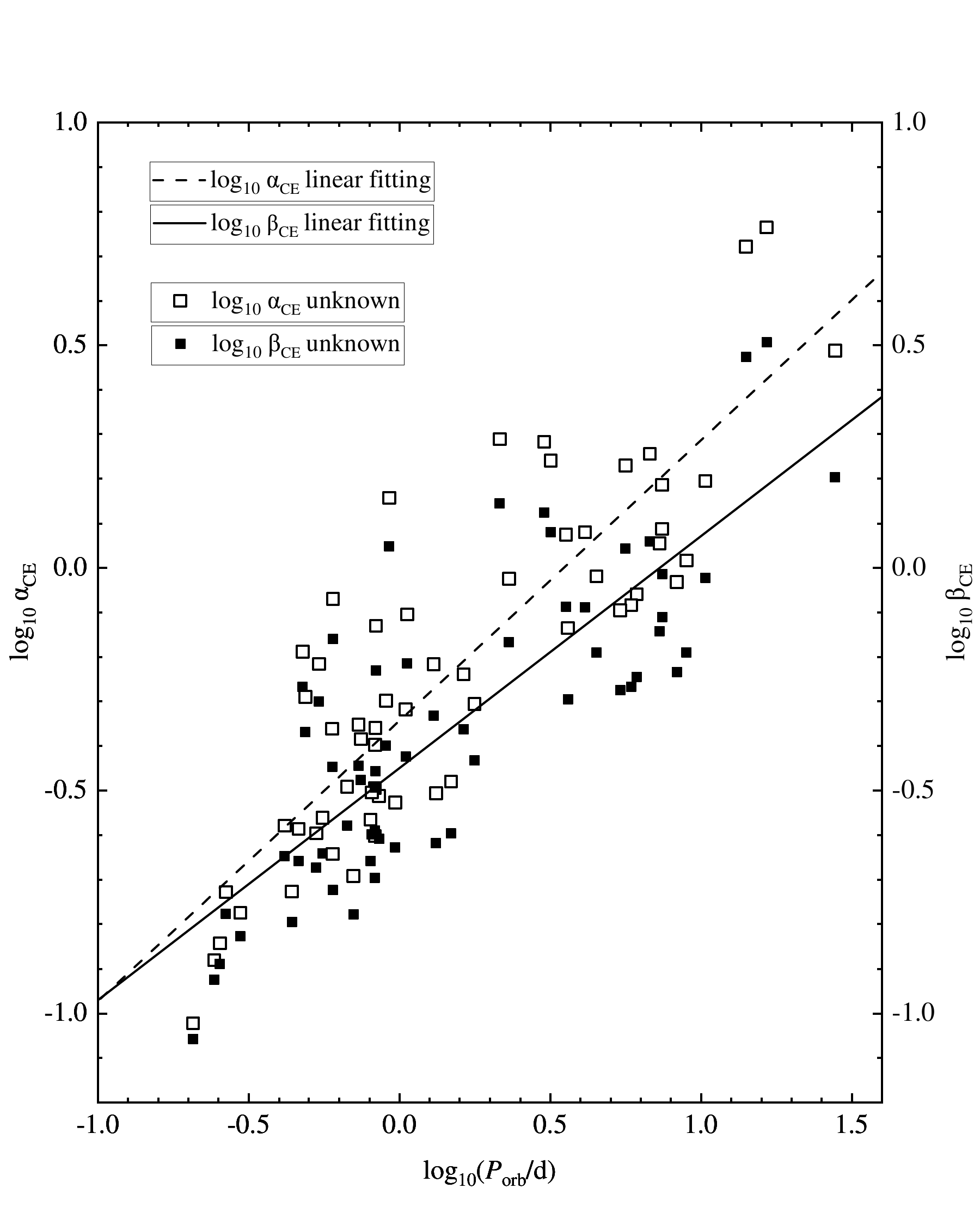}
	\caption{Similiar to Fig.\,\ref{fig05} but for sdBs with an unknown companion.
		\label{fig07}}
\end{figure}

\begin{figure}[ht!]
	\centering
	\includegraphics[scale=0.3]{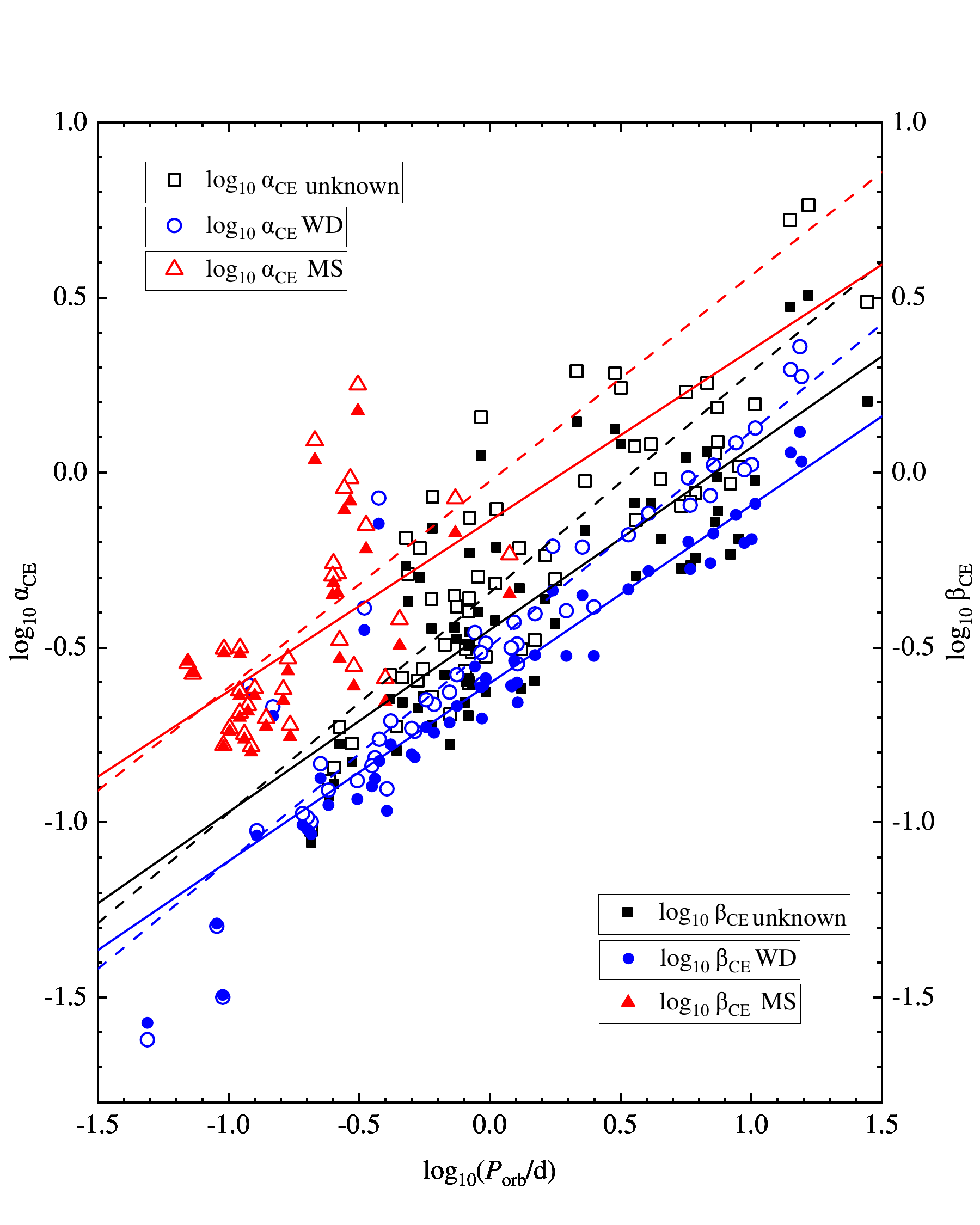}
	\caption{All data in Figs \ref{fig05} to \ref{fig07} combined. The larger CE efficiency parameter for sdB with MS companions than that for WD companions is because of the smaller companion mass (see Fig.\,\ref{fig03}).
		\label{fig08}}
\end{figure}

According to the initial mass function \citep[e.g.][]{2010A&A...520A..86Z} lower-mass stars dominate the progenitors of sdBs. So we take a $1.0\,M_\odot$ TRGB donor star as the typical progenitor of a sdB. The initial mass and radius of the donor are known. The final orbital period after CE ejection is found from the observed sdB binary. How the total energy changes as a function of remnant mass can be calculated through our adiabatic mass-loss code. With an initial guess and bisection method to solve the equations (\ref{eq07} to \ref{eq11} of this paper and 1 of \citealt{2015A&A...576A..44K}) we can find the CE parameter $\beta_{\mathrm{CE}}$ that satisfies the sdB mass $M_\mathrm{sdB}$, minimum companion mass $M_\mathrm{2}$ (assuming $i = 90^\circ$) and the orbital period $P_\mathrm{orb}$. For comparison, we also calculate $E^{'}_\mathrm{bind}$ in place of equation (\ref{eq06}) and solve for the CE parameter $\alpha_{\mathrm{CE}}$.

The solid line in Fig.\,\ref{fig04} demonstrates how the radius responds to mass loss when we assume the CE envelope is ejected adiabatically. The color of the solid line shows the binding energy as a function of the star's remnant radius $R_\mathrm{1f}$ and mass $M_\mathrm{1f}$. The dashed and dotted lines show how the inner Roche-lobe radius $R_\mathrm{L}$ changes for companion masses of $0.4\,M_\odot$ and $0.1\,M_\odot$. The black and gray colors refer to the common envelope efficiency parameters of 1.0 and 0.1. As discussed in Section\,\ref{sec:2}, the symbols are located where the remnant stars just shrink back inside their Roche lobes. Lines and symbols in Fig.\,\ref{fig04b} have the same meaning. For a specified progenitor with a given mass and evolutionary stage, the binding energy is not constant (the blue line in Fig.\,\ref{fig04b} is for $\lambda = 0.458$) as people often assumed. The binding energy varies with both the efficiency parameter $\beta_{\mathrm{CE}}$ and the companion mass $M_2$ (Fig.\,\ref{fig04b}), as well as the rebalancing of the remnant star after its envelope mass is ejected. The difference between binding energies $E_\mathrm{bind}$ and $E^{'}_\mathrm{bind}$  becomes larger for longer orbital period sdBs, and higher companion masses if the CE efficiency parameters $\alpha_\mathrm{CE}$ and $\beta_\mathrm{CE}$ are the same. Consequently, although the final mass is nearly the same as the helium core mass $M_\mathrm{He}$, defined by where mass fraction of hydrogen is 0.1 (pentagram in Fig.\,\ref{fig04}), the final radii differ from each other by orders of magnitude (see symbols in Fig.\,\ref{fig04}). The remnant core masses are all slightly larger than the helium core mass. Although the hydrogen envelope is very thin, the surface hydrogen and helium mass fractions are still similar to its progenitor's surface. 

Figs\,\ref{fig05} to \ref{fig08} present the CE efficiency parameters  $\alpha_\mathrm{CE}$ (open symbols) and $\beta_\mathrm{CE}$ (filled symbols) as functions of the orbital period of observed sdBs, of which the companions are MS, WD or unknown (MS/WD) stars. The progenitor donor is a  $1.0\,M_\odot$ TRGB star and the companion mass $M_2$ is solved with an inclination angle of $i = 90^\circ$. The dashed and solid lines are linear least-squares fits to the data (see Table \ref{tab02}). Figs\,\ref{fig05} to \ref{fig08} indicate that the CE efficiency parameters $\log_{10} \alpha_\mathrm{CE}$ and $\log_{10} \beta_\mathrm{CE}$ change linearly with the logarithm of the final orbital period if we assume the progenitor was the same. For short orbital period sdBs the CE efficiency parameters $\log_{10} \alpha_\mathrm{CE}$ and $\log_{10} \beta_\mathrm{CE}$ are similar. However a shallower slope is found for $\log_{10} \beta_\mathrm{CE}$ than for $\log_{10} \alpha_\mathrm{CE}$ in all cases for long orbital period sdBs. This agree with our expectation from Fig.\,\ref{fig04b} because the same progenitor $E_\mathrm{bind}$ gets much smaller than $E^{'}_\mathrm{bind}$ for longer orbital period sdBs. Hence, $\beta_\mathrm{CE}$ becomes less than $\alpha_\mathrm{CE}$ at longer orbital periods of sdBs. The binding energy $E_\mathrm{bind}$ from our calculation also becomes important for these long orbital period sdBs.

\subsection{1.0, 1.3 and 1.6 $M_\odot$ TRGB Donor Star}

\begin{figure}[ht!]
	\centering
	\includegraphics[scale=0.3]{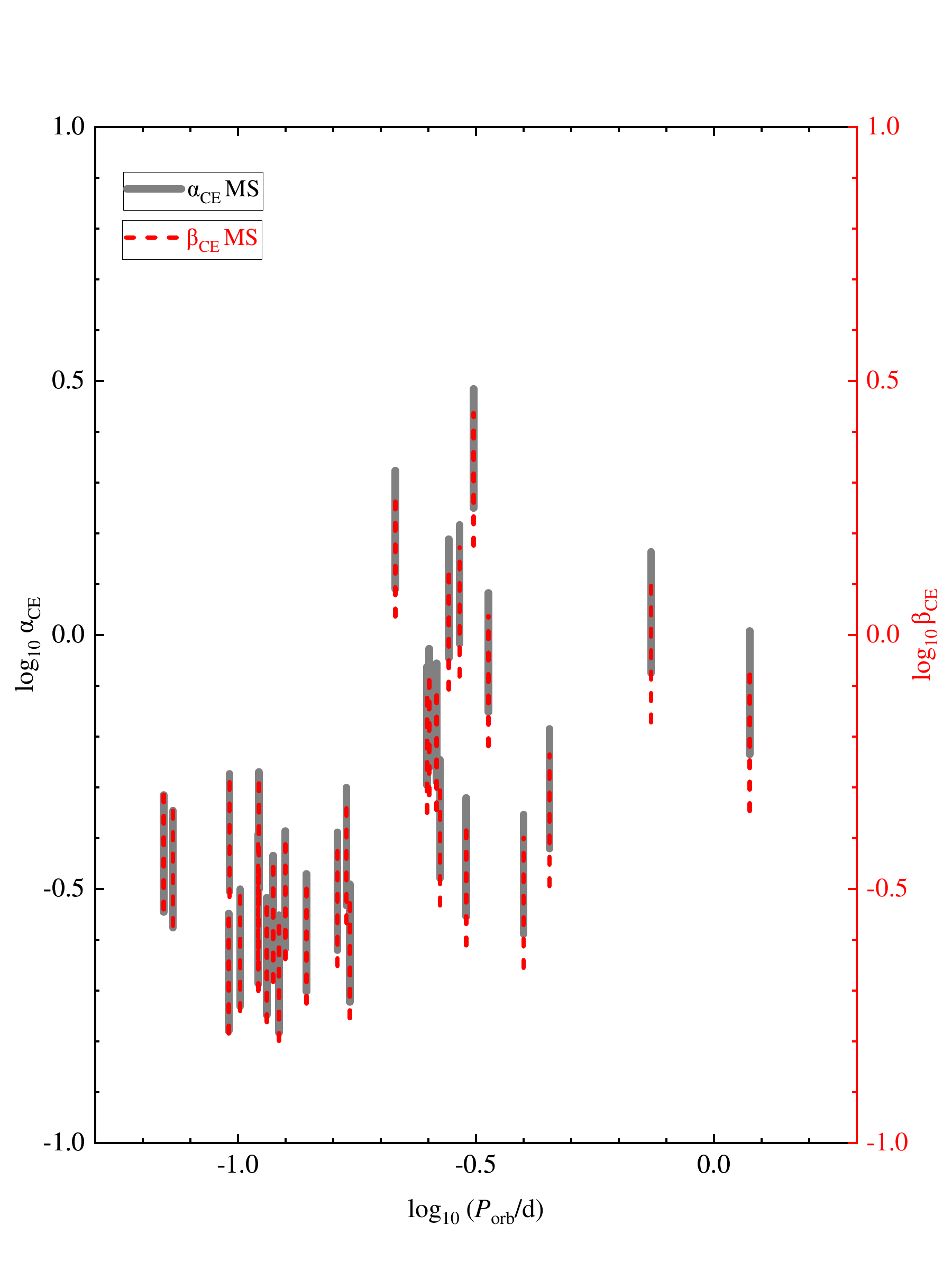}
	\caption{The CE efficiency parameters $\alpha_\mathrm{CE}$ (gray solid line) and $\beta_\mathrm{CE}$ (red dashed line) as functions of the orbital period of observed sdBs with MS companions. The progenitor's initial mass changes from $1.0M_\odot$ (bottom) to $1.6M_\odot$ (top) for each line.
	\label{fig09}}
\end{figure}

\begin{figure}[ht!]
	\centering
	\includegraphics[scale=0.3]{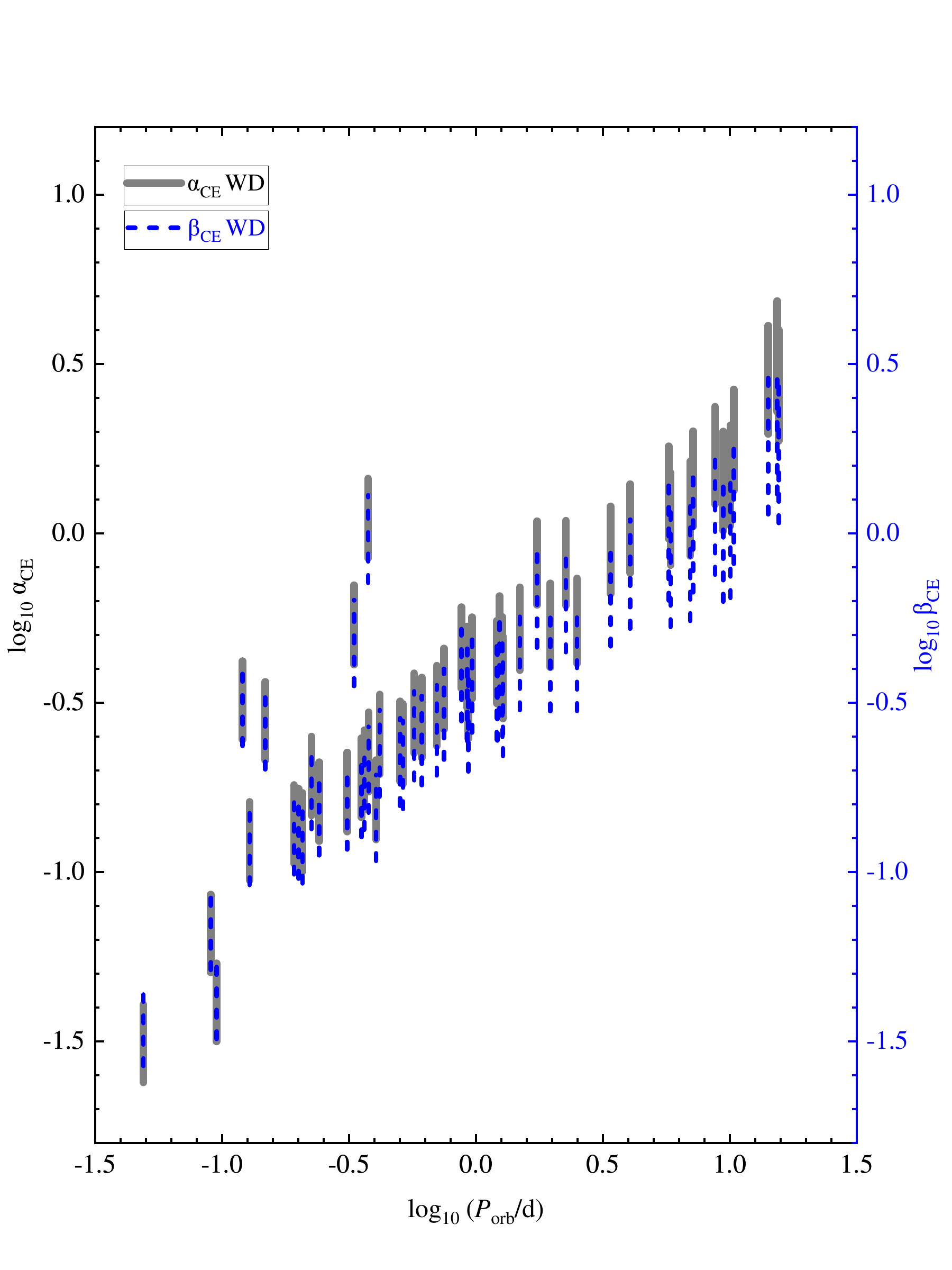}
	\caption{The CE efficiency parameters $\alpha_\mathrm{CE}$ (gray solid line) and $\beta_\mathrm{CE}$ (blue dashed line) as functions of the orbital period of observed sdBs with WD companions.
	\label{fig10}}
\end{figure}

\begin{figure}[ht!]
	\centering
	\includegraphics[scale=0.3]{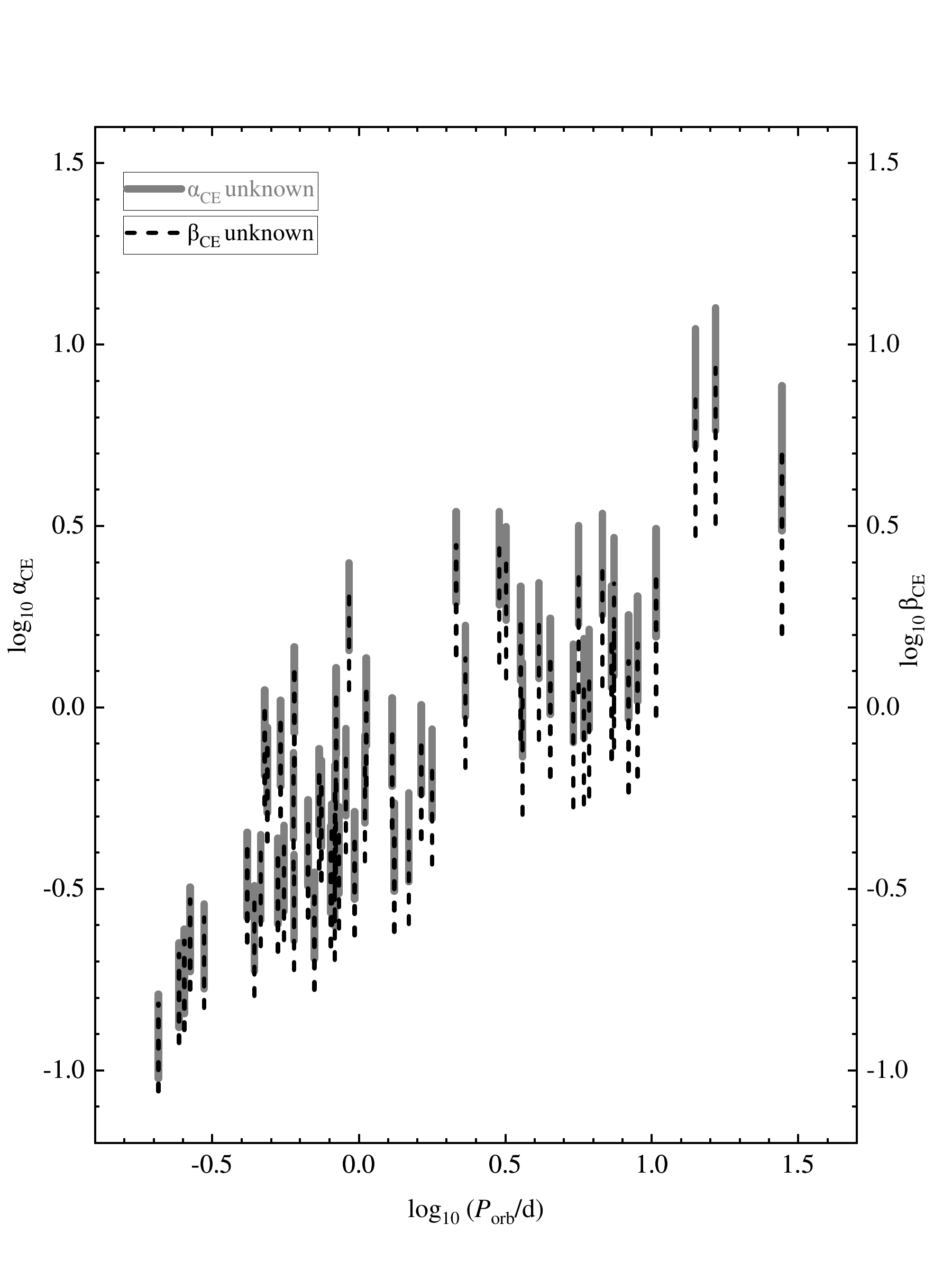}
	\caption{The CE efficiency parameters $\alpha_\mathrm{CE}$ (gray solid line) and $\beta_\mathrm{CE}$ (black dashed line) as functions of the orbital period of observed sdBs with unknown type companions.
	\label{fig11}}
\end{figure}

\begin{figure}[ht!]
	\centering
	\includegraphics[scale=0.3]{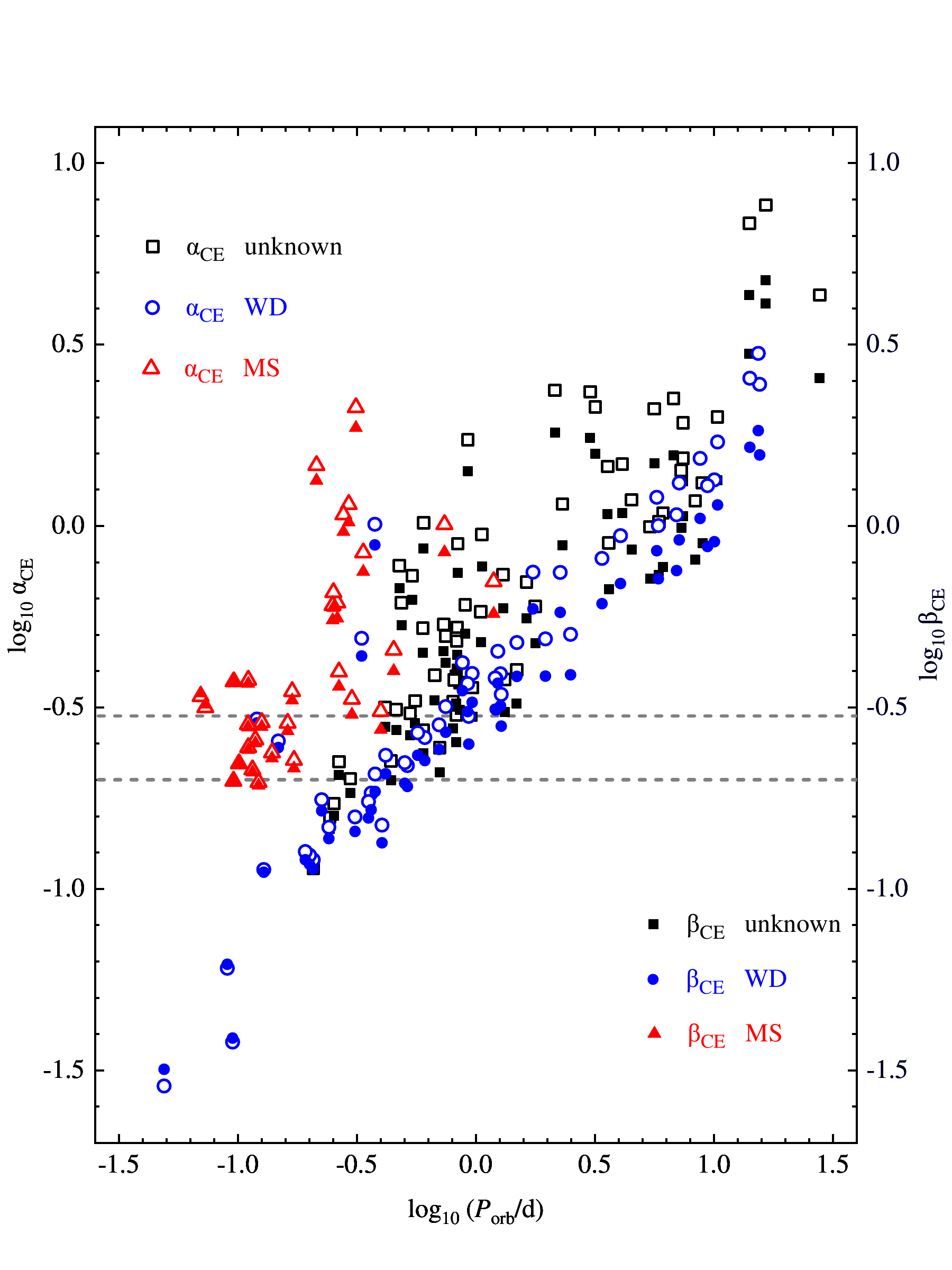}
	\caption{Weighted CE efficiency parameters $\alpha_\mathrm{CE}$ (open symbols) and $\beta_\mathrm{CE}$ (filled symbols) as functions of the orbital period of observed sdBs. Squares, circles and triangles show the unknown type, WD and MS companions. Two horizontal thin-dashed lines are CE efficiency parameters of $0.2$ and $0.3$.
	\label{fig12}}
\end{figure}

We extend our studies on the CE efficiency parameters for TRGB stars to different mass, $1.3M_\odot$ and $1.6M_\odot$, progenitors. Figs \ref{fig09} to \ref{fig11} present both the CE parameters $\alpha_\mathrm{CE}$ (solid lines) and $\beta_\mathrm{CE}$ (dashed line) as functions of the orbital periods of observed sdBs with MS, WD and unknown companions. For each line, the progenitor's initial mass changes from $1.0M_\odot$ (bottom) to $1.6M_\odot$ (top). Fig\,\ref{fig12} presents the weighted CE parameter $\alpha_\mathrm{CE}$ (open symbols) and $\beta_\mathrm{CE}$ (filled symbols). The weight factor is $ N (M_\mathrm{1i}, a_\mathrm{i}) = M^{-2.7}_\mathrm{1i}/a_\mathrm{i}$ according to the initial mass function, flat distribution of initial mass ratio and flat distribution of logarithmic initial separation \citep[see][]{2014MNRAS.444.3209H}. We find the difference between $\alpha_\mathrm{CE}$ and $\beta_\mathrm{CE}$ becomes important for sdBs with orbital periods $P_\mathrm{orb}$ such that $\log_{10} (P_\mathrm{orb}/\mathrm{d}) > -0.5$. For different progenitor masses of short orbital period sdB binaries, the binding energies $E_\mathrm{bind}$ and  $E^{'}_\mathrm{bind}$ differ significantly for $\log_{10} (P_\mathrm{orb}/\mathrm{d}) > -0.5$ with any type of companion. A higher companion mass spirals-in and ejects the envelope with a larger final radius and mass, so a smaller binding energy $E_\mathrm{bind}$ leads to a smaller $\beta_\mathrm{CE}$.

\subsection{Influence of the Binding Energy}

\begin{figure}[ht!]
	\centering
	\includegraphics[scale=0.3]{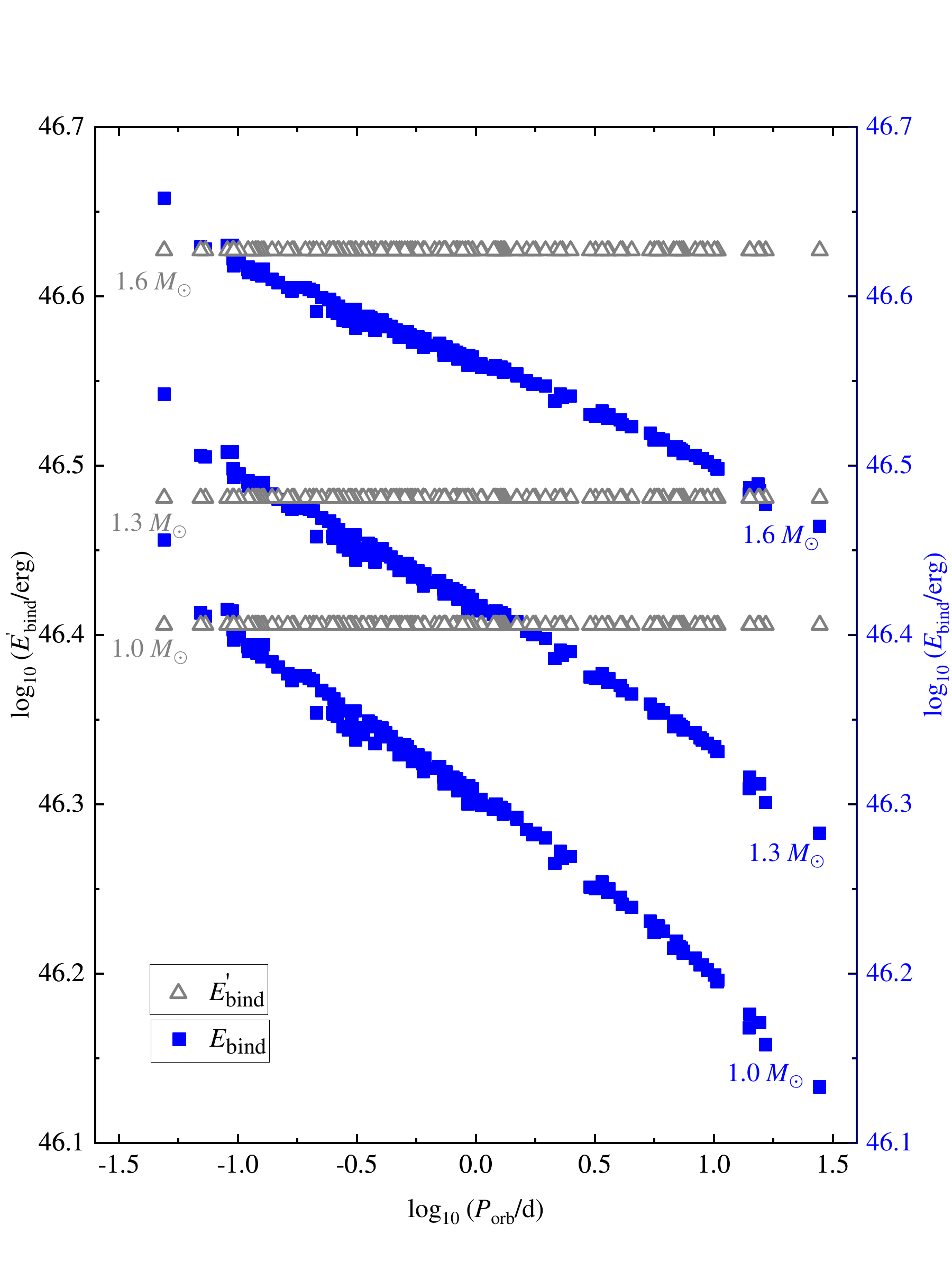}
	\caption{Binding energy as a function of orbital period of sdBs. Gray open triangles show $E^{'}_\mathrm{bind}$ as the integral of gravitational and internal energies from the progenitor's core to surface. Blue filled squares indicate the binding energy $E_\mathrm{bind}$ containing total energy, gravitational plus internal, from our adiabatic mass loss model. $E_\mathrm{bind}$ represents both the rebalance of the remnant after mass ejection and the companion's effect.
	\label{fig13}}
\end{figure}

We assume the outcome of CEE is determined by energy conservation and when both components shrink within their Roche lobes. These last two subsections provide a better understanding of the effects of different binding energy calculation methods. For accuracy, in the standard prescription of the energy formula, we calculate the binding energy $E^{'}_\mathrm{bind}$ from equation \ref{eq08} (see gray open triangles in Fig.\,\ref{fig13}). The companion mass and the final orbital period affect the binding energy and subsequently the gradient of the common envelope effciency parameter (Figs\,\ref{fig05} to \ref{fig08}, gradient not shown in Figs\,\ref{fig09} to \ref{fig12}). A flatter slope (solid lines) is seen for the new binding energy algorithm than that from a constant $E^{'}_\mathrm{bind}$ in Figs\,\ref{fig05} to \ref{fig08} and is independent of the companion type. 

For a better understanding of the difference between $\alpha_\mathrm{CE}$ and $\beta_\mathrm{CE}$, we show the binding energy $E_\mathrm{bind}$ and $E^{'}_\mathrm{bind}$ as a function the orbital period for the 142 sdBs  in Fig\,\ref{fig13}. The binding energy $E_\mathrm{bind}$ is calculated consistently as the envelope is ejected. For a given initial binary system, a shorter final orbital period requires the companion to spiral in more to eject the CE. In such cases the binding energy $E_\mathrm{bind}$ is close to $E^{'}_\mathrm{bind}$ because the the final mass is close to the Helium core mass. The binding energy $E_\mathrm{bind}$ is smaller for a longer orbital period sdBs (blue filled squares in Fig.\,\ref{fig13}) because the companions don't need to spiral in that much. $E_\mathrm{bind}$ can decrease from  $E^{'}_\mathrm{bind} = 2.548\times10^{46}$ to $1.358\times10^{46}$ erg for a $1.0M_\odot$ TRGB progenitor.

As shown in equation (\ref{eq07}), a different binding energy corresponds to a different CE efficiency. For an observed sdB system with a given progenitor model, the relation between binding energy and CE parameter is positively linearly correlated. A smaller binding energy $E_\mathrm{bind}$ reduces the CE efficiency parameter $\beta_\mathrm{CE}$ more than $\alpha_\mathrm{CE}$ for $E^{'}_\mathrm{bind}$ in long orbital period sdBs (Figs\,\ref{fig05} to \ref{fig13}).

\subsection{Relation between CE Parameter and Mass Ratio}

\begin{figure}[ht!]
	\centering
	\includegraphics[scale=0.3]{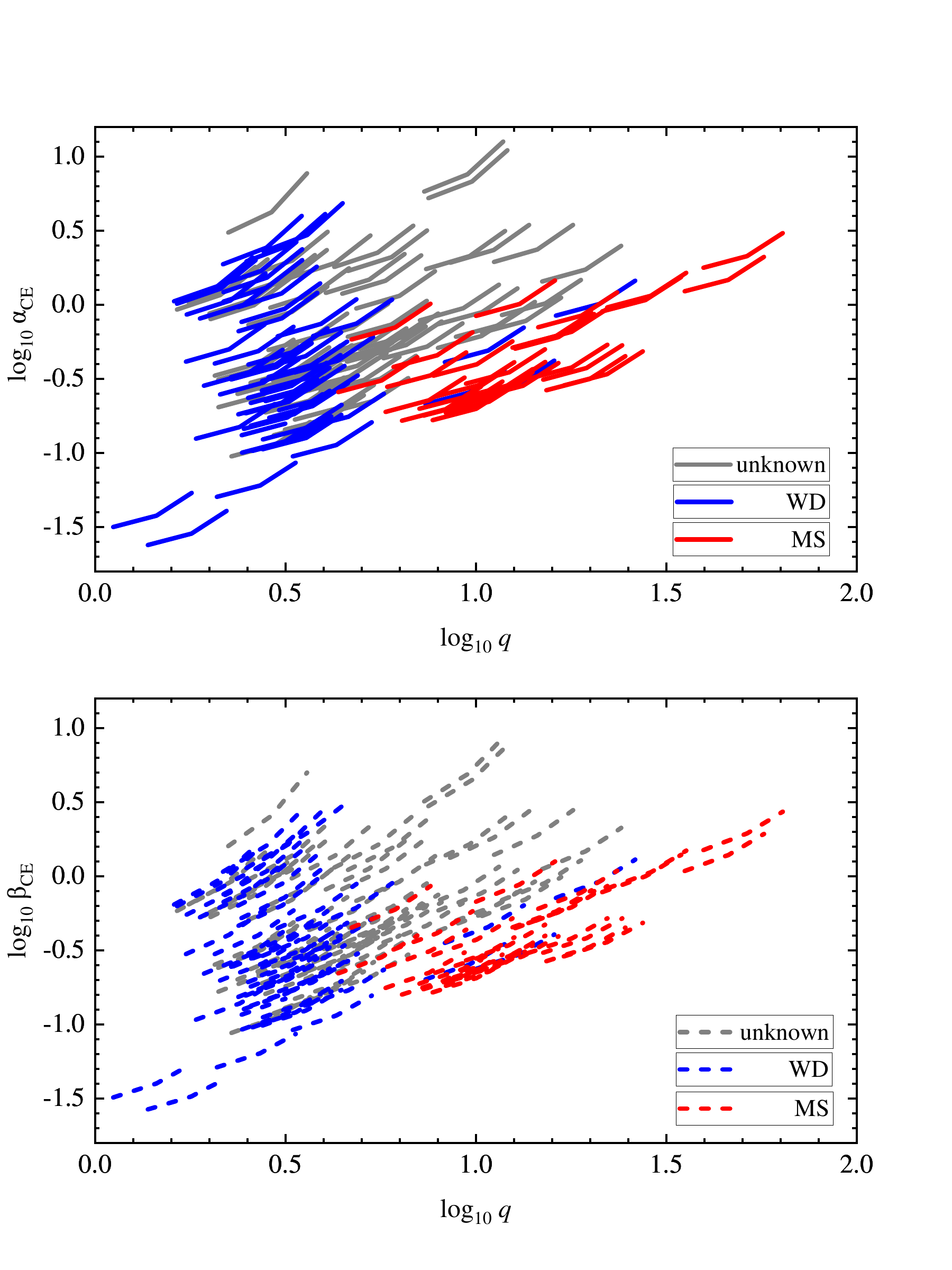}
	\caption{CE efficiency parameters as a function of mass ratio $q = M_\mathrm{1i}/ M_\mathrm{2}$. The upper panel corresponds to the binding energy from standard prescription while that of the lower panel from total energy changes. A similiar $\log 1/q$ versus $\log \alpha_{\mathrm{CE}}$ anticorrelation for post-asymptotic giant branch binaries was explored by \citet{2011MNRAS.411.2277D}. 
		\label{fig14}}
\end{figure}

If we reformulate equation (\ref{eq05}), we have the mass ratio ($q = M_\mathrm{1i}/ M_\mathrm{2}$) and CE efficiency relation,
\begin{equation}
\alpha_\mathrm{CE}  = \frac{2q(M_\mathrm{1i}-M_\mathrm{1f})}{\lambda r_\mathrm{L}(q)(M_\mathrm{1f}a_\mathrm{i}/a_\mathrm{f}-M_\mathrm{1i})} .
\label{eq12}
\end{equation}
Fig.\,\ref{fig14} shows the initial mass ratios $\log_{10} q$ to CE efficiency $\log_{10} \beta_\mathrm{CE}$ and $\log_{10} \alpha_\mathrm{CE}$ relations for short orbital period sdBs. The relation in the top panel is derived from $E^{'}_\mathrm{bind}$, while that in the bottom panel is derived from $E_\mathrm{bind}$. We find that there is linear correlation between $\log_{10} q$ and $\log_{10} \alpha_{\mathrm{CE}}$ for a given sdB system with different donor masses. Although the relation is stretched for different sdB systems, the linear correlation is universal, sharing the same gradient. We have $\log_{10} \alpha_\mathrm{CE} = -0.503 (\pm 0.049) + 0.328 (\pm 0.062) \times \log_{10} q$ with the ajusted R-squared of 0.059 and $\log_{10} \beta_\mathrm{CE} = -0.631 (\pm 0.044) + 0.400 (\pm 0.055) \times \log_{10} q$ with the adjusted R-squared of 0.108.

\citet{2011MNRAS.411.2277D} explored such an anticorrelation between $\log 1/q$ and $\log \alpha_{\mathrm{CE}}$ for post-asymptotic giant branch (post-AGB) binaries. They found the gradient to be around $-1.2$, but they excluded post-RGB donors. The results for sdBs in this paper also show the $\log 1/q$ versus $\log \alpha_{\mathrm{CE}}$ anticorrelation (see equation \ref{eq12}). 

\subsection{Effects from Other Sources}

\begin{figure}[ht!]
	\centering
	\includegraphics[scale=0.231]{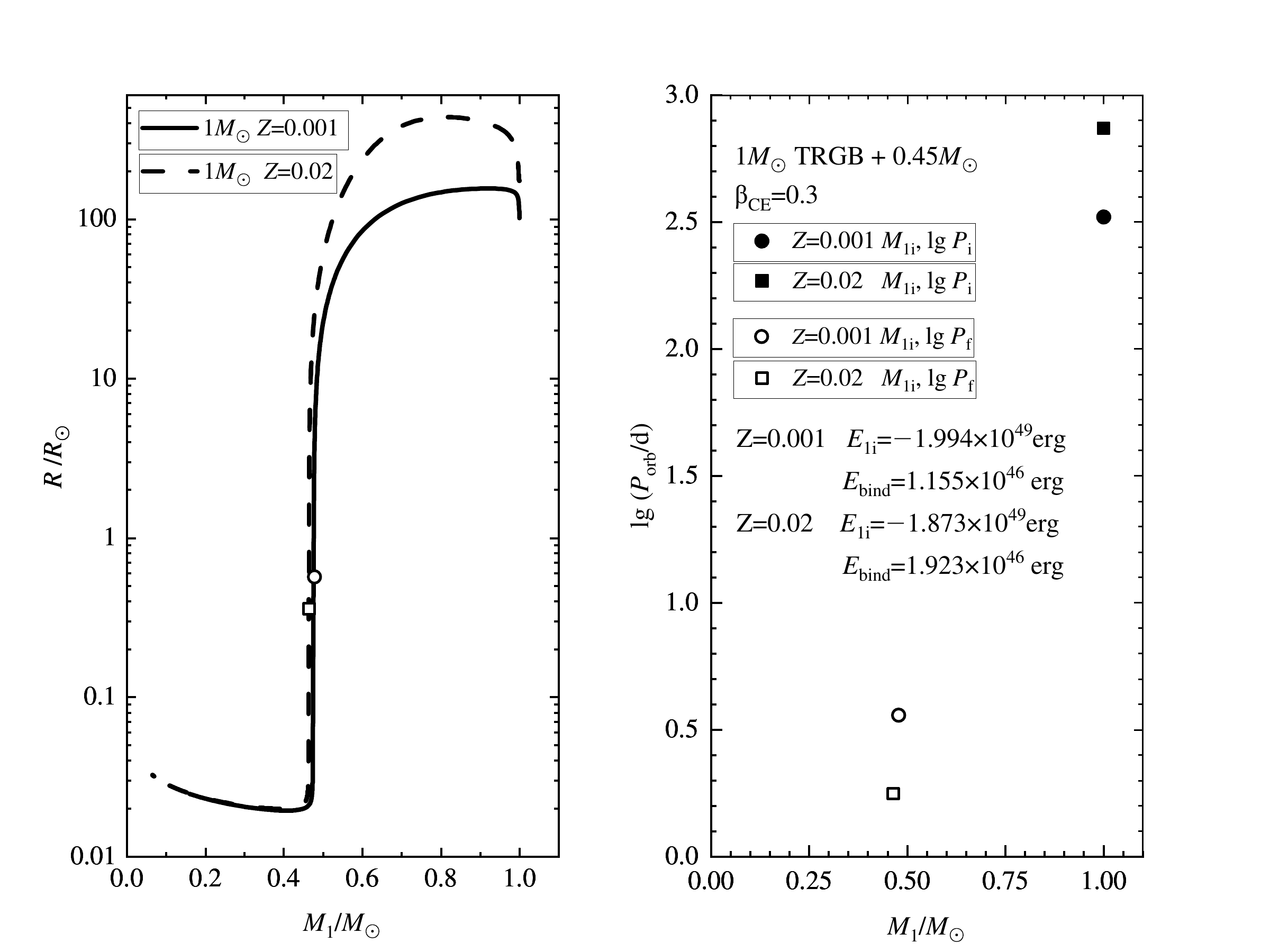}
	\caption{Metallicity dependence of the radius response (left panel) and the initial-final mass and orbital period (right panel). The solid line and dashed line show the radius response of stars with $Z=0.001$ and $Z=0.02$. The solid and open symbols (squares for  $Z=0.02$ and dots for $Z=0.001$) are the initial and final mass and orbital period. Metal poor stars are more compact and can shrink to their Roche lobes earlier than metal rich stars. 
		\label{fig15}}
\end{figure}

To constrain CE efficiency parameters, we use a new binding energy algorithm addressing the influence of the companion mass and the rebalance of the remnant core. The progenitors of short orbital period sdBs are located at a narrow range near the tip of RGB. We only show the results of TRGB stars in the previous subsections. Progenitors from a slightly earlier stage of TRGB extend a little the range of CE efficiency parameters. The role that convection might play in transporting energy out, as \citet{arXiv:2203.06091} suggest, might effect CEE efficiencies. We discuss other possible effects on the uncertainty of CE efficiency parameters such as metallicity, companion type, and inclination angle in this subsection.

We only explored stars with metallicity $Z = 0.02$. Metal-poor stars are more compact than metal-rich stars with the same mass. We find that metal-poor stars have smaller radii and shorter orbital periods for the same companion mass (Fig.\,\ref{fig15}). The binding energy is highly related to the radius of the donor star. We expect that a metal-poor star has a larger binding energy $E^{'}_\mathrm{bind}$ than a metal-rich star. For a given CE efficiency parameter $\beta_{\mathrm{CE}} = 0.3$, the final orbital period or separation is surprisingly larger for a metal-poor star (see equation\,\ref{eq07}). This is because the compact metal-poor star shrinks to its Roche lobe earlier and makes $E_\mathrm{bind}$ smaller. So a smaller metallicity tends to reduce the linear slope of the logarithm of the CE efficiency parameter and final orbital period. Unfortunately it is hard to spectrally determine the metallicity of short orbital period hot subdwarf binaries (private comm with Prof. Yangping Luo). For example, the companions of sdBs are all cool stars with spectral type around $M$ \citep{2015A&A...576A..44K} while the sdB star is too luminous, unlike the long orbital period sdBs which indicate the chemical evolution history of Milk Way \citep{2020A&A...641A.163V}. We currently cannot pick out the observed sdBs with metallicity $Z = 0.02$ to give a stricter prediction or constraint.

\begin{figure}[ht!]
	\centering
	\includegraphics[scale=0.231]{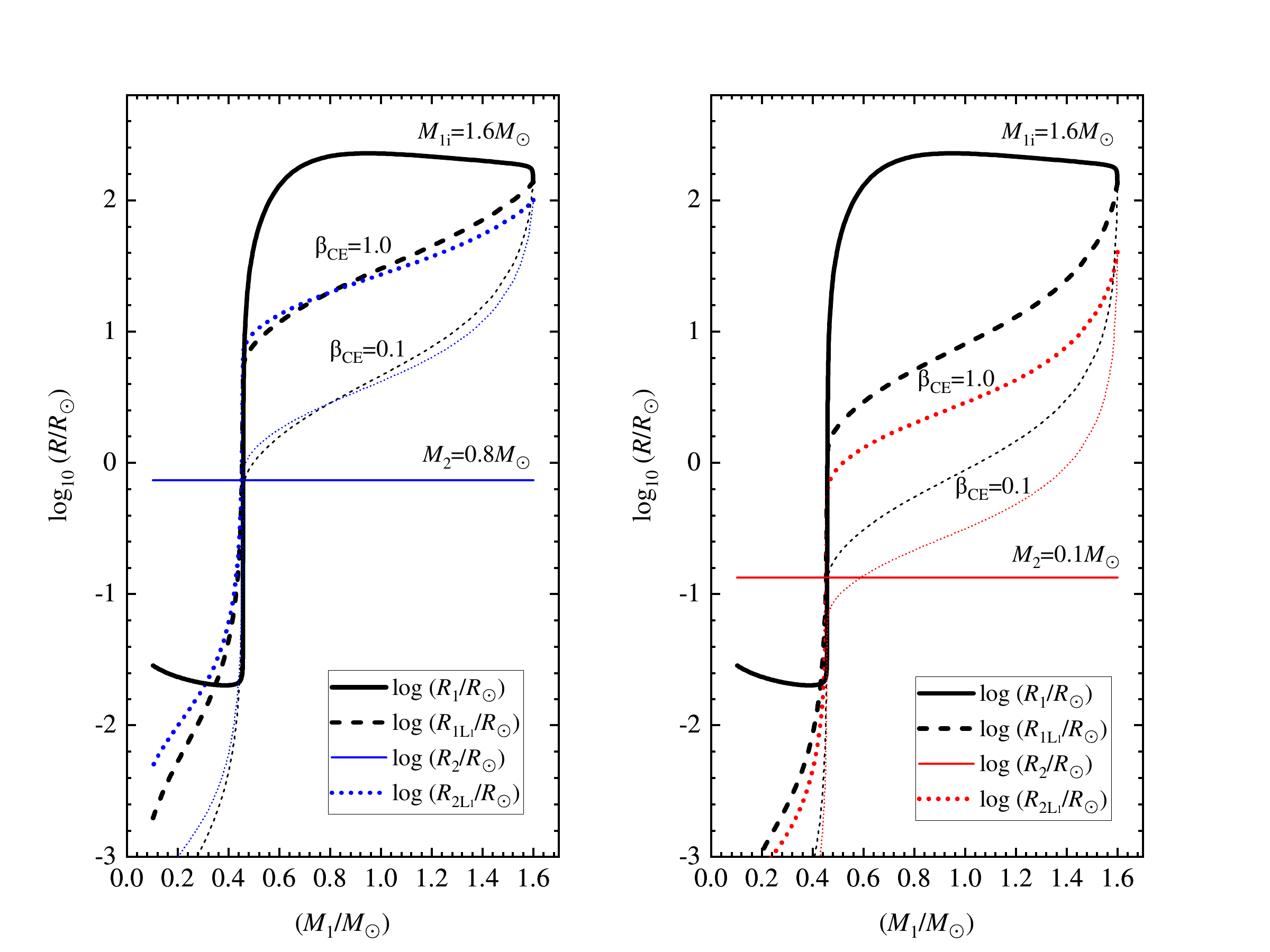}
	\caption{The remnant radius and Roche-lobe radius as a function of mass. Left and right panels are a $1.6\,M_\odot$ TRGB star with a $0.8\,M_\odot$ and a $0.1\,M_\odot$ MS companion. The black solid line is the radius response of the donor star. Dashed and dotted lines are the Roche-lobe radius of the donor and the companion. Thick and thin lines show the CE efficiency parameters of $1.0$ and $0.1$.
		\label{fig16}}
\end{figure}

The companion mass range is different for short orbital period sdBs with a MS or WD companion (Fig.\,\ref{fig03}). From theoretical analysis, if the MS companion mass $M_2$ is too close to that of its companion $M_\mathrm{1i}$ it is likely to merge with the envelope. Because the companion's radius is likely to be larger than its Roche-lobe radius (Fig.\,\ref{fig16}). For example if $M_2$ were larger than $1.1\,M_\odot$ for a $M_\mathrm{1i}=1.6 M_\odot$ TRGB star with $\beta_\mathrm{CE}=0.1$. Furthermore, the MS companion mass cannot be too massive because at the initial mass ratio $q < q_\mathrm{ad}$, it avoids CE evolution (right panel of Fig.\,\ref{fig02}). A combination of the above reasons is very likely to explain the lower mass peak around $0.1\,M_\odot$ in Kupfer's Fig.\,8. There is also a lower limit around $0.1$ for the CE efficiency parameters of sdBs with a MS companion. A low-mass MS companion does not shrink inside its Roche-lobe if the CE efficiency parameter is too small (Fig.\,\ref{fig16}). For extreme short orbital period sdBs with a WD companion, the CE efficiency parameters are smaller than 0.1 (Figs\,\ref{fig06}, \ref{fig08}, \ref{fig10} and \ref{fig12}). The progenitors of these sdBs are possibly more massive than $2.5\,M_\odot$. They likely form through the CE + CE channel. We only explore the progenitors of sdBs with a degenerate core and our current results do not cover progenitors of sdBs with a non-degenerate core.

In this study, we assume the inclination angle $i = 90^\circ$. We might expect there is a random distribution of the inclination angle, flat in $\cos i$. However, the companion mass is degenerate with the inclination angle \citep[equation 1 of ][]{2015A&A...576A..44K}. We might hope to constrain the CE efficiency parameters through checking the random number density of sdBs with respect to inclination angle (see similiar method for milisecond pulsars with helium WD by \citealt{2014MNRAS.437.2217S}) but the typical companion mass of sdBs is more extended than the neutron star mass in their study. We hope future observations can provide inclination angles and metallicities in short orbital period binaries with a compact companion.

We might not provide a concluding answer for whether there is a universal constant CE efficiency \citep{2020cee..book.....I} $\alpha_\mathrm{CE}$ but we are trying to provide comprehensive studies and discussions on this aspect. Our results tend to support that longer orbital periods sdBs are from lower mass progenitors while shorter orbital periods sdBs are from higher mass progenitors (Figs.\,\ref{fig09} to \ref{fig10}). If the above assumption is true we can connect the top part of the leftmost dashed line and the bottom part of the rightmost dashed line in these Figures. Consequently, we can find a constant $\log_{10} \beta_\mathrm{CE} \approxeq -0.3$ for sdBs with a MS companion and from $\log_{10} \beta_\mathrm{CE} \approxeq -0.8$ to  $\log_{10} \beta_\mathrm{CE} \approxeq -0.2$ for sdBs with a WD companion (abandoning the 3 leftmost objects might from non-degenerate progenitors and 3 rightmost objects with orbital period longer than 10 days in Fig.\,\ref{fig10}). In other words, the full orbital period range of sdBs can be covered within that CE efficiency parameter range. A concluding answer needs detailed binary population synthesis studies addressing the full range of the progenitors, metallicity, effects in this study, and more strict observation as discussed.

\section{Summary and Conclusions}
\label{sec:5}

In this paper we calculate a binding energy $E_\mathrm{bind}$ that differs from previous studies by tracing the change of the total energy based on our adiabatic mass loss model. For comparison, we also calculate the normal binding energy $E^{'}_\mathrm{bind}$ through integrating the total energy from the core to the surface. We assume the outcome of CEE is determined by energy conservation and when both components shrink within their Roche lobes. We have found that the remnant core radius can vary by orders of magnitude although the remnant core mass is almost the same. The remnant core mass is slightly larger than the helium core. We examine short orbital period sdBs to estimate CE efficiency parameters with different binding energy calculation methods. The CE efficiency parameter is normally defined as $\alpha_\mathrm{CE}$ in the standard energy formula (\ref{eq05}). We write the CE efficiency parameter as $\beta_\mathrm{CE}$ for our binding energy method. For shorter orbital period sdBs, the binding energy is almost the same and we find the difference between $\alpha_\mathrm{CE}$ and $\beta_\mathrm{CE}$ is small. For longer orbital period sdBs in our samples, we have found that the binding energy can differ by up to a factor of 2. The CE efficiency parameter $\beta_{\mathrm{CE}}$ becomes smaller than $\alpha_\mathrm{CE}$ for final orbital periods $\log_{10} P_{\mathrm{orb}}/\mathrm{d} > -0.5$. A shallower slope for $\log_{10} \beta_{\mathrm{CE}}$ against $\log_{10} P_{\mathrm{orb}}$ than that for $\log_{10} \alpha_{\mathrm{CE}}$ is found for sdBs with all types of companion. We also find a $\log q$ and $\log \alpha_{\mathrm{CE}}$ linear correlation in sdBs similar to what \citet{2011MNRAS.411.2277D} found in the post-AGB binaries.
	
To constrain the CEE physics more precisely, \citet{2014MNRAS.444.3209H} improved the core radius fit through detailed stellar model calculations. \citet{2021A&A...648L...6P} improved the radius at the mass-radius flat range \citep{1996ApJ...458..692T} by studying planetary nebulae with binary cores.  \citet{2011MNRAS.411.2277D,2011ApJ...743...49L} provide useful constraints on the efficiency parameter $\alpha_\mathrm{CE}$ and structure parameter $\lambda$ or the binding energy as function of the stellar mass and evolutionary stage. However, our approach differs from earlier studies. We calculate the new binding energy $E_\mathrm{bind}$ by tracing the difference between the initial and final total energy of the donor star so that the influence of the companion mass on the binding energy is addressed. For more massive close binaries, such as those with black holes or neutron stars, that come from more massive progenitors than sdBs, the difference between $E_\mathrm{bind}$ and $E^{'}_\mathrm{bind}$ is probably more important. Effects such as launching of jets may also play an important role \citep{2022RAA....22e5010S}.

Different studies constraining the outcome of CEE process, such as our method, are important for binary population synthesis studies. More pre- and post-CE binary observations with more accurate orbital and structure parameters, as well as metallicity, are also of benefit. We shall present the grid of the CE efficiency parameters and other remnant information with different donor masses and evolutionary stages in the future. 

Acknowledgments

The authors thank the anonymous referee for the helpful comments and suggested improvements. This project is supported by National Key R\&D Program of China (2021YFA1600400/3) and National Natural Science Foundation of China (grants NO. 12173081, 12090040/3, 11733008, 12125303, 11673058), Yunnan Fundamental Research Projects (grant NO. 202101AV070001), the key research program of frontier sciences, CAS, No. ZDBS-LY-7005 and CAS, “Light of West China Program”. HG thanks the institute of astronomy, University of Cambridge for hosting the one year visiting. HG also thanks Prof. Ronald F Webbink for helpful suggestions. CAT thanks Churchill Colledge for his fellowship. DJ acknowledge the science research grants NO. 12073070 and CMS-CSST-2021-A08. Zhenwei Li acknowledges the science research grants No. 12103086, 202101AU070276.


\bibliography{hongwei_sdB_manuscript}{}
\bibliographystyle{aasjournal}

\end{document}